\documentclass{article}
\usepackage{iclr2027_conference,times}
\usepackage[T1]{fontenc}
\usepackage[utf8]{inputenc}
\usepackage{amsmath,amssymb,graphicx,booktabs,longtable,array,calc}
\usepackage{fancyvrb}
\usepackage{hyperref}
\usepackage{xurl}

\hypersetup{
  hidelinks,
  pdfauthor={Hongye Yang, Zhihao Xie, Shengjun Xiong, Boxiao Huang},
  pdftitle={DepthBenchCAD: When Does Deeper Auditing Yield More Reliable Conclusions?}
}

\title{DepthBenchCAD: When Does Deeper Auditing Yield More Reliable Conclusions?}

\author{
Hongye Yang \\
College of Computing \\
Georgia Institute of Technology \\
\texttt{hyang783@gatech.edu}
\And
Zhihao Xie \\
Independent Researcher \\
\texttt{xiezhihao.ai@gmail.com}
\AND
Shengjun Xiong \\
Independent Researcher \\
\texttt{xiongshengjunchina@gmail.com}
\And
Boxiao Huang \\
College of Computing \\
Georgia Institute of Technology \\
\texttt{bhuang361@gatech.edu}
}

\iclrfinalcopy

\begin{document}
\maketitle
\lhead{Preprint}

\begin{abstract}

Generative CAD models are expected to remain behaviorally correct after
parameter edits, so increasing the number of edit checks is often
treated as a direct route to more reliable evaluation. Under a fixed
budget, however, auditing each program more thoroughly reduces the
number of tasks and independent generations that can be evaluated, which
can ultimately make model-level estimates less accurate. We study this
phenomenon and the conditions under which it arises. We decompose
behavioral evaluation into three evidence levels---task templates,
stochastic generations, and within-program edits---define an average
failure risk that is invariant to audit depth, and combine three-level
variance with measured execution costs to analyze the tradeoff between
deeper edit auditing and broader independent coverage. Experiments
across two CAD environments and five generation systems show that the
value of deeper auditing depends on where evaluation uncertainty
originates. When template heterogeneity or generation stochasticity
dominates, additional edit checks can increase total estimation error;
when within-program state variation is large and generation is
expensive, deeper auditing is more valuable. Variance and cost estimates
from calibration predict the direction of this change and provide a
diagnostic basis for allocating evidence on held-out tasks. These
results show that the thoroughness of program inspection can diverge
from the reliability of model evaluation, and they help determine
whether the next unit of budget should be spent on a new task, a new
generation, or additional edit checks.

\end{abstract}


\section{Introduction}\label{introduction}

Outputs from parametric CAD models must generate correct geometry at
default parameters and remain executable while preserving geometric
constraints and design intent under subsequent edits. DeepCAD represents
the CAD modeling process as an operation sequence \citep{wu2021},
while Fusion 360 Gallery provides real design histories and programmatic
construction data \citep{willis2021}. Text2CAD maps natural language
to parametric CAD sequences \citep{khan2024}, and CAD-Recode uses
language models to recover executable CAD code from geometric input
\citep{rukhovich2025}. A program may appear correct in its initial
state yet fail to build, self-intersect, invalidate parameters, or
violate constraints after changes to thickness, hole diameter, spacing,
or array count. Static geometric metrics therefore do not fully capture
the editability or engineering utility of CAD programs.

Behavioral evaluation of generative CAD contains three levels of
evidence: task templates cover different part families and design
difficulty; independent generations from the same template capture model
stochasticity; and counterfactual edit audits characterize behavior
across valid parameter states within a program. BenchCAD incorporates
execution verification, parameter reasoning, and code editing into a
programmatic CAD benchmark \citep{zhang2026}. Text2CAD-Bench further
provides systematic evaluation of text-to-parametric-CAD generation
\citep{wang2026}. Increasing evidence at any of the three levels can
reduce evaluation uncertainty, but the statistical role and execution
cost of each level differ. Under a fixed budget, excessive audit depth
reduces task coverage, whereas shallow auditing may miss conditional
failures. The numbers of templates, generations per template, and audits
per program must therefore be chosen jointly.

To ensure that different audit depths evaluate the same capability, we
freeze the valid edit states for each template in advance and define the
model-level estimand as the average probability of behavioral failure
when randomly sampling a template, one model generation, and one edit
state. Audit depth changes only the precision with which this estimand
is measured, allowing budget allocations to be compared directly.

Supplementary motivation and protocol explanation appear in Appendix~\ref{supplementary-explanation}. Our main contributions are:

(1) We identify and empirically characterize a separation between the
thoroughness of program inspection and the accuracy of model-level
estimation, showing how deeper auditing under a fixed budget can
increase estimation error by reducing task coverage or independent
generations.

(2) We explain this effect through three-level variance and execution
cost and test, on held-out data, whether these quantities predict the
direction of the benefit from deeper auditing.

(3) We evaluate how evidence allocation affects model-evaluation
reliability through risk estimation, interval coverage, and judgment
error, while explicitly reporting the associated calibration cost.

\section{Related Work}\label{related-work}
Executable CAD generation retains design histories and construction logic through operation sequences, programmatic data, and recovered code \citep{wu2021,willis2021,khan2024,rukhovich2025}. BenchCAD and Text2CAD-Bench extend evaluation toward execution, parameter reasoning, and editing \citep{zhang2026,wang2026}. Our analysis concerns how to allocate evaluation evidence under a fixed budget. It builds on multistage sampling \citep{neyman1934,horvitz1952}, clustered observations \citep{liang1986}, and work on stochastic evaluation and selection bias \citep{henderson2018,reimers2017,dietterich1998,varma2006}. Detailed CAD, representation-learning, judge-validation, and sampling literature appears in Appendix~\ref{extended-related-work}.

\section{Method}\label{method}

To characterize how much independent evidence each evaluation record
actually provides, we model generative-CAD behavioral evaluation as a
three-level nested sampling process over templates, generations, and
edits. This section defines the estimand, sampling estimator, variance
decomposition, cost constraint, and calibration strategy. Letting the
design structure determine effective sample size follows basic
principles of design-based sampling and finite-population inference
\citep{neyman1934,horvitz1952}.

\subsection{Evaluation Target and Finite Edit Population}\label{evaluation-target-and-finite-edit-population}

Let \(Y_{tim}^{(s)} \in \left\{ 0,1 \right\}\) denote the outcome for
system \(s\) on template \(t\) for generation \(i\) evaluated at state
\(m\), where \(Y_{tim}^{(s)} = 1\) indicates an execution error,
timeout, invalid topology, geometric-constraint violation, or parameter
response inconsistent with the task specification. The failure risk of a
single program over the complete edit population is

\begin{equation}
\mu_{s}(t,i) = \frac{1}{M}\sum_{m = 1}^{M}Y_{tim}^{(s)}.
\label{eq:1}
\end{equation}

The model-level estimand is defined as

\begin{equation}
R_{s} = \mathbb{E}_{T}\mathbb{E}_{I^{(s)} \mid T}\left\lbrack \mu_{s}\left( T,I^{(s)} \right) \right\rbrack.
\label{eq:2}
\end{equation}

\(R_{s}\) is the average probability of behavioral failure when randomly
sampling a task template, a model generation, and a valid edit state.
Actual audit depth affects estimation cost and variance but does not
change this estimand. Single-program certification, all-state pass rate,
and worst-case risk are different evaluation targets and are outside the
scope of this paper.

\subsection{Three-Level Sampling and Risk Estimation}\label{three-level-sampling-and-risk-estimation}

Given a formal evaluation budget, we first sample \(q\) templates from
the available pool; then independently generate \(g\) programs for each
template; finally, from each program\textquotesingle s \(M\) frozen
states, we sample \(k\) states without replacement. Let
\(\mathcal{S}_{ti}\) denote the set of sampled states for program
\((t,i)\); the model-risk estimator is

\begin{equation}
{\widehat{R}}_{s}(q,g,k) = \frac{1}{q}\sum_{t = 1}^{q}\frac{1}{g}\sum_{i = 1}^{g}\frac{1}{k}\sum_{m \in \mathcal{S}_{ti}}^{}Y_{tim}^{(s)}.
\label{eq:3}
\end{equation}

The three sampling levels serve different roles: \(q\) controls task
coverage; \(g\) controls generation uncertainty within a task; \(k\)
controls measurement error in an individual program\textquotesingle s
risk. Multiple generations from the same template share task difficulty,
so the template is the outermost unit for variance estimation, data
splitting, and bootstrap. Keeping the template as the outermost cluster
also avoids treating correlated repeated observations as independent
samples \citep{liang1986}.

When comparing multiple systems, all systems use the same templates and
edit-state indices to reduce comparison noise from task difficulty. A
template and all of its generated programs are always kept as an intact
cluster and never split across calibration, validation, or test sets.

\subsection{Three-Level Variance Decomposition}\label{three-level-variance-decomposition}

Let
\(\sigma_{T}^{2} = {Var}_{T}\left\lbrack \mathbb{E}_{I \mid T}\mu_{s}(T,I) \right\rbrack\)
denote between-template variance,
\(\sigma_{G}^{2} = \mathbb{E}_{T}\left\lbrack {{Var}_{I \mid T}\mu}_{s}(T,I) \right\rbrack\)
within-template generation variance, and
\(\sigma_{E}^{2} = \mathbb{E}_{T,I}\left\lbrack S_{E}^{2}\left( Y_{tim}^{(s)} \right) \right\rbrack\)
within-program edit-state variance, respectively. Under a balanced
sampling design, the superpopulation variance of Equation~\eqref{eq:3} is

\begin{equation}
{Var}\left( {\widehat{R}}_{s} \right) = \frac{\sigma_{T}^{2}}{q} + \frac{\sigma_{G}^{2}}{qg} + \frac{1}{qg}\left( \frac{1}{k} - \frac{1}{M} \right)\sigma_{E}^{2}.
\label{eq:4}
\end{equation}

Equation~\eqref{eq:4} shows that increasing audit depth \(k\) can reduce only
edit-state sampling error; between-template variation and generation
stochasticity still require increasing \(q\) and \(g\). For a finite
reference pool with \(Q\) templates, \(G\) complete generations per
template, and \(M\) states per program, resampling replay uses the
finite-population variance in Equation~\eqref{eq:5}. Estimation without
replacement and inclusion-probability methods provide the classical
basis for this resampling design on a frozen reference pool \citep{horvitz1952}.

\begin{equation}
{Var}_{ref}\left( {\widehat{R}}_{s} \right) = \left( \frac{1}{q} - \frac{1}{Q} \right)S_{T}^{2} + \frac{1}{q}\left( \frac{1}{g} - \frac{1}{G} \right)S_{G}^{2} + \frac{1}{qg}\left( \frac{1}{k} - \frac{1}{M} \right)S_{E}^{2}.
\label{eq:5}
\end{equation}

Equation~\eqref{eq:5} is used to validate the method on a finite test pool.
Confidence intervals for the future task distribution use templates as
bootstrap clusters and do not treat the finite template set as the
complete task population. We use the bootstrap to estimate uncertainty
for the future task distribution; its statistical basis and
confidence-interval properties are well established \citep{efron1979,efron1986}.

\subsection{Cost Model and Joint Allocation}\label{cost-model-and-joint-allocation}

Let \(c_{T}\) be the preparation and scheduling cost of introducing a
new template, \(c_{I}\) be the cost of one model generation, parsing,
and initial build, and \(c_{E}\) be the cost of one parameter injection,
program re-execution, and automated judgment. The total cost of a
balanced design is

\begin{equation}
C(q,g,k) = qc_{T} + qg\left( c_{I} + kc_{E} \right).
\label{eq:6}
\end{equation}

Because systems have different generation and execution costs, our
shared strategy freezes the number of generations per template \(g\) and
audit depth per program \(k\), together with the budget-mapping rule,
and measures the cost parameters for system \(s\) before formal
evaluation. The maximum number of templates available to that system
within the budget is

\begin{equation}
q_{s}(g,k) = \min\left\{ Q,\left\lfloor \frac{C_{0}}{c_{T,s} + g\left( c_{I,s} + kc_{E,s} \right)} \right\rfloor \right\}.
\label{eq:7}
\end{equation}

Thus, different systems may use different numbers of templates \(q_{s}\)
while sharing the same \(g\), \(k\), and budget-allocation rule.

Normalizing by the cost of one edit execution \(c_{E}\) and defining
\(r_{T} = c_{T}/c_{E}\), \(r_{I} = c_{I}/c_{E}\), the budget constraint
becomes

\begin{equation}
qr_{T} + qg\left( r_{I} + k \right) \leq C_{0}.
\label{eq:8}
\end{equation}

We use discrete search to jointly select the three sampling parameters:

\begin{equation}
\arg{\min_{q,g,k \in \mathbb{N}}{{\widehat{Var}}_{s}(q,g,k)}},\quad\text{s.t.}\quad C(q,g,k) \leq C_{0}.
\label{eq:9}
\end{equation}

The search range satisfies \(1 \leq q \leq Q\),\(1 \leq g \leq G\) and
\(1 \leq k \leq M\). For each \((g,k)\), the algorithm chooses the
largest number of templates \(q\) allowed by the budget and then
compares all feasible integer solutions. This procedure handles template
saturation, finite-population corrections, and residual integer budget.

\subsection{Calibration Cost and Usage Strategy}\label{calibration-cost-and-usage-strategy}

Calibration uses independent templates to estimate variance and cost. A pooled $(g,k)$ configuration requires only target-system cost measurement; system-level calibration diagnoses deviations and reuse scenarios. Appendices~\ref{a.3-cost-constraint-continuous-approximation-and-integer-search}, \ref{f.4-formal-selection-contract-for-allocation-strategies}, and \ref{supplementary-explanation} give the allocation and calibration details.

\section{Experimental Design}\label{experimental-design}

\subsection{Data and Evaluation Environments}\label{data-and-evaluation-environments}

We compare five generation systems, S1--S5, on DepthBenchCAD. All five
use a common CAD-generation interface and differ only in the underlying
language model; each system receives the same task specification and
parameter requirements and outputs an executable CadQuery program.
System mappings, fixed runtime configurations, random-seed rules, and
cost-measurement protocols are provided in Appendix~\ref{appendix-d.-system-mapping-runtime-environment-and-frozen-protocol}.

The experiments use a primary environment A and a cross-task-family
transfer environment B. Environment A contains eight task families and
is used to test whether calibration parameters predict the benefit of
deeper auditing on held-out templates. Environment B contains six task
families disjoint from A and is used to test cross-family transfer of
allocation strategies. The environments share the same risk definition
and four edit protocols; complete task-family compositions and execution
settings appear in Appendices \ref{appendix-b.-task-families-and-construction-of-frozen-edit-states} and \ref{appendix-d.-system-mapping-runtime-environment-and-frozen-protocol}.

\begingroup
\setlength{\tabcolsep}{3pt}
\begin{longtable}[]{@{}
  >{\raggedright\arraybackslash}p{(\columnwidth - 14\tabcolsep) * \real{0.2541}}
  >{\raggedright\arraybackslash}p{(\columnwidth - 14\tabcolsep) * \real{0.0786}}
  >{\raggedright\arraybackslash}p{(\columnwidth - 14\tabcolsep) * \real{0.0875}}
  >{\raggedright\arraybackslash}p{(\columnwidth - 14\tabcolsep) * \real{0.1216}}
  >{\raggedright\arraybackslash}p{(\columnwidth - 14\tabcolsep) * \real{0.1187}}
  >{\raggedright\arraybackslash}p{(\columnwidth - 14\tabcolsep) * \real{0.1464}}
  >{\raggedright\arraybackslash}p{(\columnwidth - 14\tabcolsep) * \real{0.0774}}
  >{\raggedright\arraybackslash}p{(\columnwidth - 14\tabcolsep) * \real{0.1157}}@{}}
\caption{Experimental data matrix}\label{tab:1}\\
\toprule\noalign{}
\textbf{Env.} & \textbf{Tpl.} & \textbf{Fam.} & \textbf{Cal./Test} & \textbf{Gen./tpl.} & \textbf{States/prog.} & \textbf{Sys.} & \textbf{Records} \\
\midrule\noalign{}
\endfirsthead
\toprule
\textbf{Env.} & \textbf{Tpl.} & \textbf{Fam.} & \textbf{Cal./Test} & \textbf{Gen./tpl.} & \textbf{States/prog.} & \textbf{Sys.} & \textbf{Records} \\
\midrule\noalign{}
\endhead
\bottomrule\noalign{}
\endlastfoot
DepthBenchCAD-A & 72 & 8 & 24/48 & 5 & 16 & 5 & 28,800 \\
DepthBenchCAD-B & 48 & 6 & 12/36 & 4 & 16 & 5 & 15,360 \\
\end{longtable}
\endgroup

Using one edit execution as the cost unit, the template cost is 2.
Generation costs for S1--S5 are 1, 3, 8, 16, and 32 in environment A,
and 2, 6, 16, 32, and 64 in environment B. The main reported budget is
\(C_0\)=1024, with total cost q{[}2+g(rI+k){]}.

The main experimental results use a frozen, real, fully audited pool.
Failure labels come from actual builds, counterfactual edits, and
automated judgments. Alternative evidence allocations are evaluated only
by sampling without replacement from observed records and by Monte Carlo
replay. Complete execution, cost, and replay protocols are provided in
Appendices \ref{appendix-c.-isolated-execution-and-automated-judge}, \ref{appendix-d.-system-mapping-runtime-environment-and-frozen-protocol}, and F.

\subsection{Counterfactual Edit States and Automated Judge}\label{counterfactual-edit-states-and-automated-judge}

Each template contains 16 pre-frozen edit states, evenly divided among
four categories.

\begingroup
\setlength{\tabcolsep}{3pt}
\begin{longtable}[]{@{}
  >{\raggedright\arraybackslash}p{(\columnwidth - 6\tabcolsep) * \real{0.1408}}
  >{\raggedright\arraybackslash}p{(\columnwidth - 6\tabcolsep) * \real{0.0929}}
  >{\raggedright\arraybackslash}p{(\columnwidth - 6\tabcolsep) * \real{0.3199}}
  >{\raggedright\arraybackslash}p{(\columnwidth - 6\tabcolsep) * \real{0.4464}}@{}}
\caption{Composition of counterfactual edit states}\label{tab:2}\\
\toprule\noalign{}
\textbf{Type} & \textbf{States} & \textbf{Intervention target} & \textbf{Primary checks} \\
\midrule\noalign{}
\endfirsthead
\toprule
\textbf{Type} & \textbf{States} & \textbf{Intervention target} & \textbf{Primary checks} \\
\midrule\noalign{}
\endhead
\bottomrule\noalign{}
\endlastfoot
Local edit & 4 & Routine change to one parameter & Parameter binding,
geometric response, and main-body connectivity \\
Boundary edit & 4 & Parameter values near the valid-range boundary &
Self-intersection, zero volume, minimum spacing, and manufacturing
constraints \\
Linked edit & 4 & Simultaneous changes to multiple related parameters &
Symmetry, spacing, arrays, and dimensional relations \\
Semantic edit & 4 & High-level design-specification changes &
Coordinated response of related parameters, geometry, and design
semantics \\
\end{longtable}
\endgroup

The state generator reads only the task specification and never the
candidate program implementation. Each state records parameter values,
units, valid ranges, expected geometric relations, and judgment
tolerances. Complete definitions of the 16 states are provided in
Appendix~\ref{appendix-b.-task-families-and-construction-of-frozen-edit-states}.

The automated judge checks, in order, successful build completion, valid
entities and connected components, topology and hole/circular features,
parameter response, dimensional/spacing/symmetry/containment relations,
and high-level semantic constraints. The full contract for numerical
tolerances, gray-zone routing, isolated execution, and geometry probes
appears in Appendix~\ref{appendix-c.-isolated-execution-and-automated-judge}.

If the initial state fails to build or violates nominal constraints, all
16 edit states for that program are counted as failures in the primary
analysis. We also report conditional risk among initially valid programs
to distinguish initial-generation failure from failures introduced
during editing.

\paragraph{Expert validation.}

The automated judge is validated on 800 stratified records reviewed by
two experts. The experts label each record independently and blindly
before a consensus label is formed; overall risk and confusion metrics
are weighted by inverse sampling probability. Appendix \ref{appendix-e.-expert-validation} gives the full
sampling, adjudication, agreement, and weighted-estimation protocol.

\subsection{Data Splits and Calibration Setup}\label{data-splits-and-calibration-setup}

Splits are stratified by family at the template level: A uses three calibration and six test templates per family (24/48 total), and B uses two and six (12/36). All generations and states stay with their template. Full calibration audits all 16 states; smaller pilots test shrinkage and reuse. Test data never select variance parameters, shrinkage weights, depth, or costs; the full test pool supplies only reference risk and replay outcomes \citep{varma2006}. Appendices~\ref{d.3-execution-cost-and-randomness-protocol}, \ref{f.1-calibration-cost}, and \ref{supplementary-explanation} retain the extended protocol.

\subsection{Comparison Methods and Budgets}\label{comparison-methods-and-budgets}

We compare fixed audit depths, a two-level degenerate model, pooled and
leave-one-system-out pooling, full system-level calibration, small
calibration with shrinkage, edit-type stratification, and paired
allocation for model pairs. Main-text results focus on fixed depth,
pooled allocation, and system-level/stratified candidate configurations.
Formal selection rules, budget mappings, and paired definitions are
provided in Appendix~\ref{appendix-f.-supplementary-experiments-robustness-and-model-comparison}.

\subsection{Evaluation Metrics and Statistical Inference}\label{evaluation-metrics-and-statistical-inference}

\textbf{Risk-Estimation Efficiency}

We measure estimation error at both the program and model levels.
Program-level MSE is the mean squared error between the failure
proportion estimated from k sampled states and the failure proportion
over all 16 states for that program, averaged across programs.
Model-level MSE uses the mean failure risk of the full test pool as the
reference and reports J=\(C_0\)\ensuremath{\times}{}MSE.

For each fixed audit depth, only the calibration set is used to choose
g, and the largest feasible q is determined by the cost constraint.
Program-level MSE is estimated by state-subsampling replay, while
model-level MSE is computed from the design variance of the finite test
pool. Test-pool results are used only for post hoc validation and never
for configuration selection.

Benefit prediction is preregistered for two comparisons: k=4\ensuremath{\to}{}8 and
k=8\ensuremath{\to}{}16. We define \ensuremath{\Delta}{}J=\(J_{\mathrm{deep}}\)\ensuremath{-}{}\(J_{\mathrm{shallow}}\), where a negative value
indicates that deeper auditing is beneficial. We compare the sign
predicted during calibration with the actual \ensuremath{\Delta}{}J in the test pool.
Wrong-recommendation loss is the excess J of the recommended
configuration over the smaller J of the two alternatives.

In addition to J and benefit direction, we report regret relative to the
oracle, 5\% robust coverage, 95\% interval coverage, automated-judge
validity, and paired model decisions. Definitions, repetition counts,
and statistical inference for these auxiliary metrics are given in
Appendices \ref{appendix-e.-expert-validation} and \ref{appendix-f.-supplementary-experiments-robustness-and-model-comparison}. Interval experiments use q=16, g=3, k=8 throughout
and run 5,000 replays without replacement.

\section{Experimental Results}\label{experimental-results}

This section addresses three questions: whether ignoring evidence
hierarchy produces overconfident model conclusions; whether three-level
variance and cost predict the direction of benefit from deeper auditing;
and when recalibration is worth its additional cost.

\subsection{Evidence Hierarchy and Evaluation Reliability}\label{evidence-hierarchy-and-evaluation-reliability}

When generated programs are treated as independent units, mean interval
coverage across the five systems is 91.81\%; using the three-level
design variance increases it to 95.26\%. The difference is largest for
S1 and S2, which exhibit stronger template-level correlation: coverage
rises from 89.06\% and 86.82\% to 94.28\% and 94.42\%, respectively.
Differences for S3--S5 are smaller, consistent with their weaker
template-level heterogeneity. These results describe interval
performance on the finite test pool and do not change the common risk
point estimate used by both methods.

\begin{figure}[tbp]
\centering
\includegraphics[width=\linewidth]{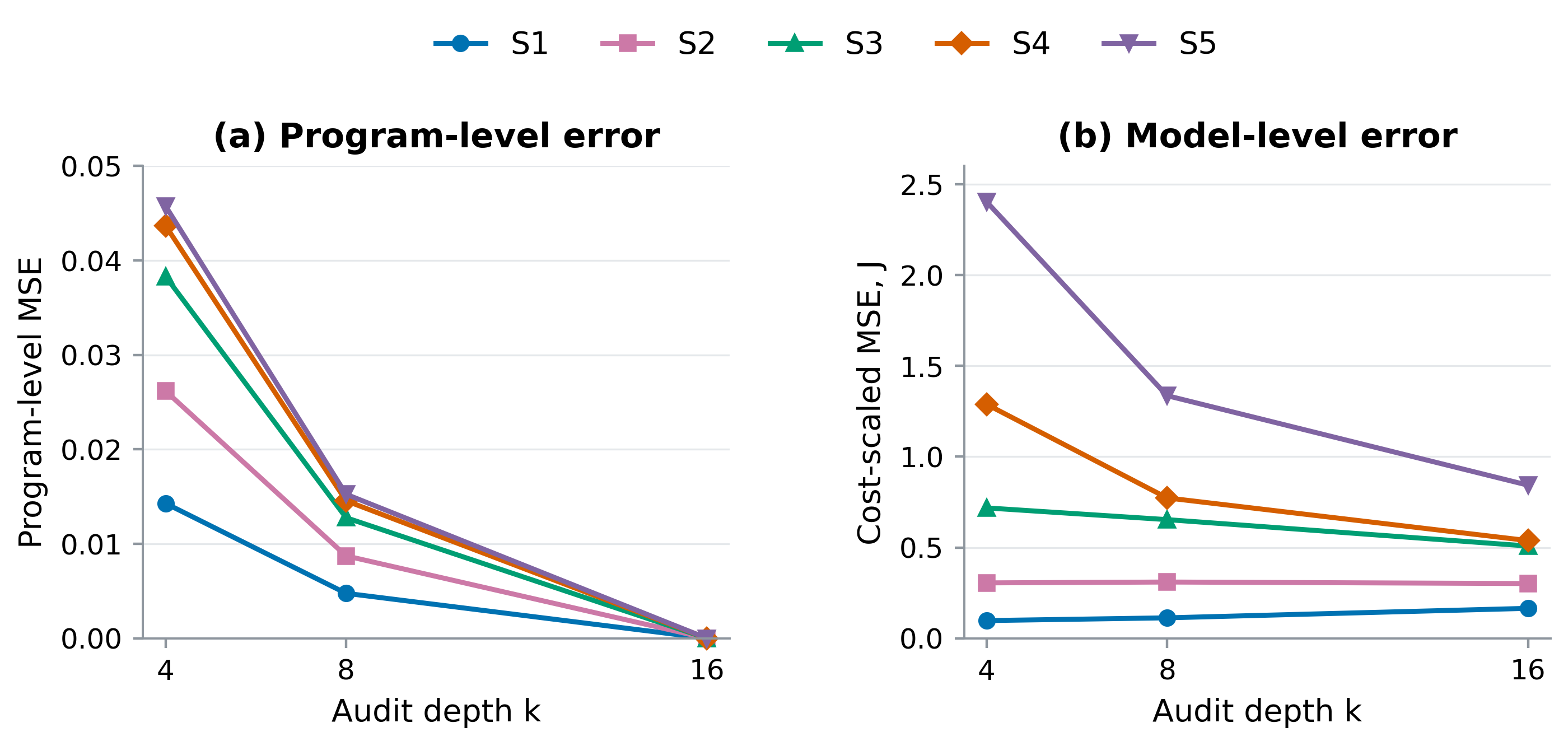}
\caption{Audit depth and estimation error in the primary environment.
(a) Program-level MSE. (b) Model-level cost-scaled error, J=\(C_0\)\ensuremath{\times}{}MSE. The
formal budget is \(C_0\)=1024; the numbers of templates q and generations g
vary by configuration. Colors and markers denote S1--S5, and lines
connect the three reported depths. At k=16, program-level MSE is zero,
while model-level error may still increase or decrease. Exact values are
reported in Table~\ref{tab:F6}.}
\label{fig:1}
\end{figure}

Figure~\ref{fig:1} compares program-level and model-level estimation errors at
different audit depths in the primary environment.

More thorough program inspection does not improve model-level estimation
accuracy for every system. When S1\textquotesingle s audit depth
increases from 4 to 16, program-level MSE falls from 0.014284 to zero,
but the total number of generated programs drops from 184 to 48 and
model-level J rises from 0.0974 to 0.1649, an increase of about 69.3\%.
Full-state auditing eliminates edit-sampling error for these programs
but cannot compensate for the loss of independent generations.

S4 shows the opposite pattern. Increasing depth from 4 to 16 reduces the
number of templates from 46 to 30, yet J falls from 1.2881 to 0.5378, a
decrease of about 58.3\%. Here, the information gained from additional
edit states is sufficient to offset the loss of coverage. S2 changes
only slightly, indicating that some systems are nearly indifferent
across audit depths.

Across the ten predefined comparisons in environment A, calibration
predicts the correct direction in nine. The sole error is S2 for k=4\ensuremath{\to}{}8:
predicted \ensuremath{\Delta}{}J=\ensuremath{-}{}0.0121, while actual \ensuremath{\Delta}{}J=0.0039. Mean wrong-recommendation
loss across the ten comparisons is 0.000389. After local calibration in
environment B, all ten directions are predicted correctly. These results
support the three-level variance-and-cost explanation of audit benefit,
although the twenty comparisons constitute only a limited mechanism
validation.

\begin{figure}[tbp]
\centering
\includegraphics[width=\linewidth]{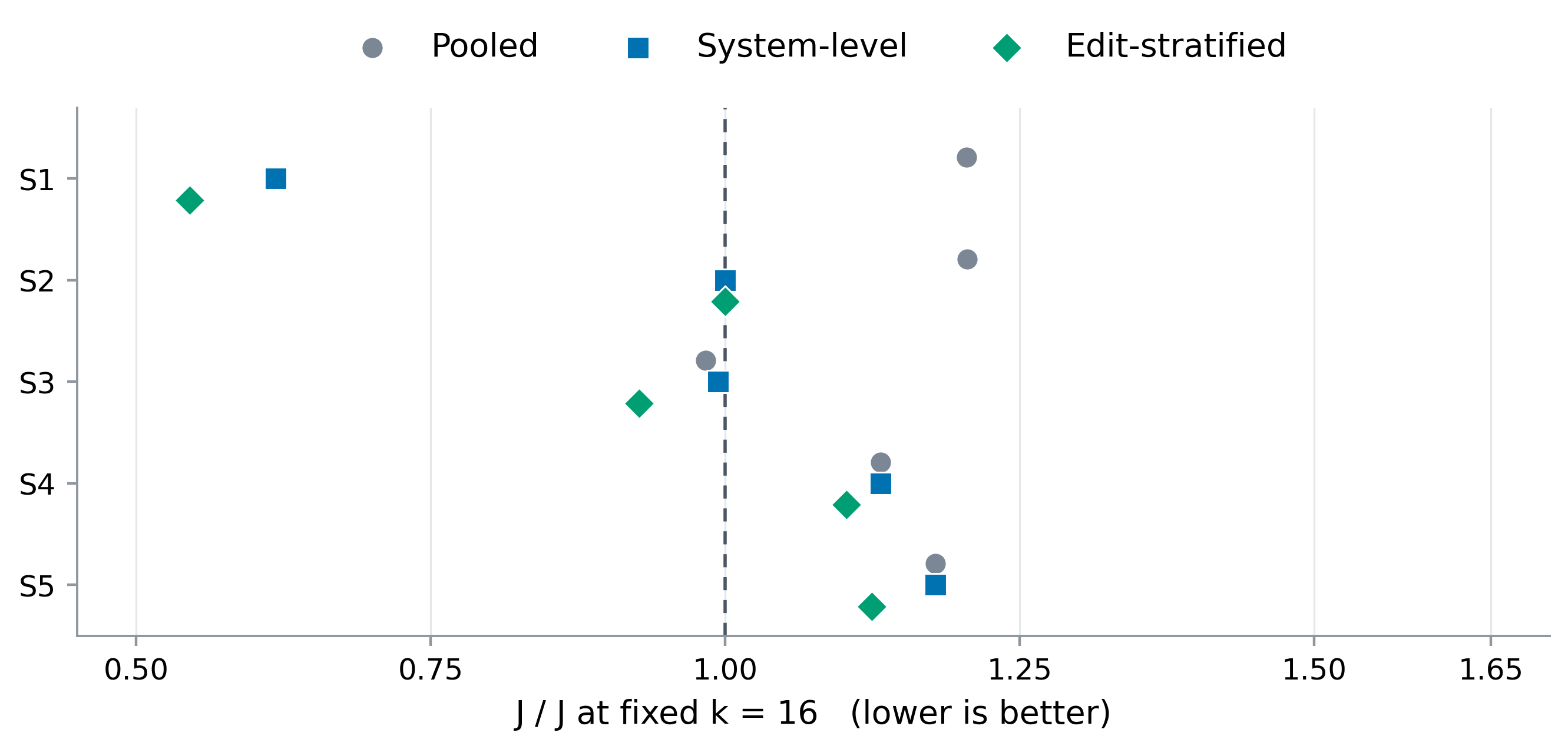}
\caption{Error ratios of three allocation strategies relative to the
fixed k=16 baseline in the primary environment. Each point is the
strategy\textquotesingle s J divided by J for the same system at fixed
k=16. The dashed line marks a ratio of 1; points to the left have lower
error. Circles, squares, and diamonds denote pooled, system-level, and
edit-stratified strategies, respectively. The formal budget is \(C_0\)=1024;
calibration cost is accounted for separately. Exact values are reported
in Table~\ref{tab:F7}.}
\label{fig:2}
\end{figure}

Figure~\ref{fig:2} shows three classes of candidate configurations selected under
a common calibration protocol. System-level and stratified
configurations reduce estimation error for some systems, but neither
pooled nor system-level configurations consistently outperform a strong
fixed-depth baseline. The three-level model therefore provides a stable
explanation of the source and direction of audit benefit, while the
optimality of a particular configuration still depends on system
variance structure, cost, and finite calibration data.

A strong fixed depth remains an important baseline. Under the unified
analysis protocol, fixed \(k = 16\), the stratified configuration, the
system-level configuration, and the cross-system pooled configuration
have five-system mean \(J\) values of 0.471, 0.481, 0.502, and 0.533. We
therefore do not interpret calibration-driven joint allocation as
outperforming the best fixed depth on average. Its primary role is to
explain why different systems prefer different audit depths and to
predict the direction of benefit from increasing audit depth.

\subsection{Calibration Cost and Transfer Stability}\label{calibration-cost-and-transfer-stability}

Calibration is substantially more expensive than a single formal
evaluation: full calibration costs about 2.04--5.67 times the formal
budget, while small calibration costs about 28.1\%--76.6\%. Any local
gain from system-level allocation must therefore be weighed against
calibration expenditure and the number of times calibration can be
reused. Complete values are given in Table~\ref{tab:F1}.

Repeated template splits reveal finite-sample variability in audit-depth
selection. Across random splits, median relative regret is 1.041 and
mean relative regret is 1.098. Tail degradation is more pronounced under
leave-one-task-family-out validation: P90 relative regret reaches 1.709
for S1 and 1.608 for S2, showing that task-family coverage affects
allocation transfer. Complete stability results appear in Table~\ref{tab:F2}.

\begin{figure}[tbp]
\centering
\includegraphics[width=\linewidth]{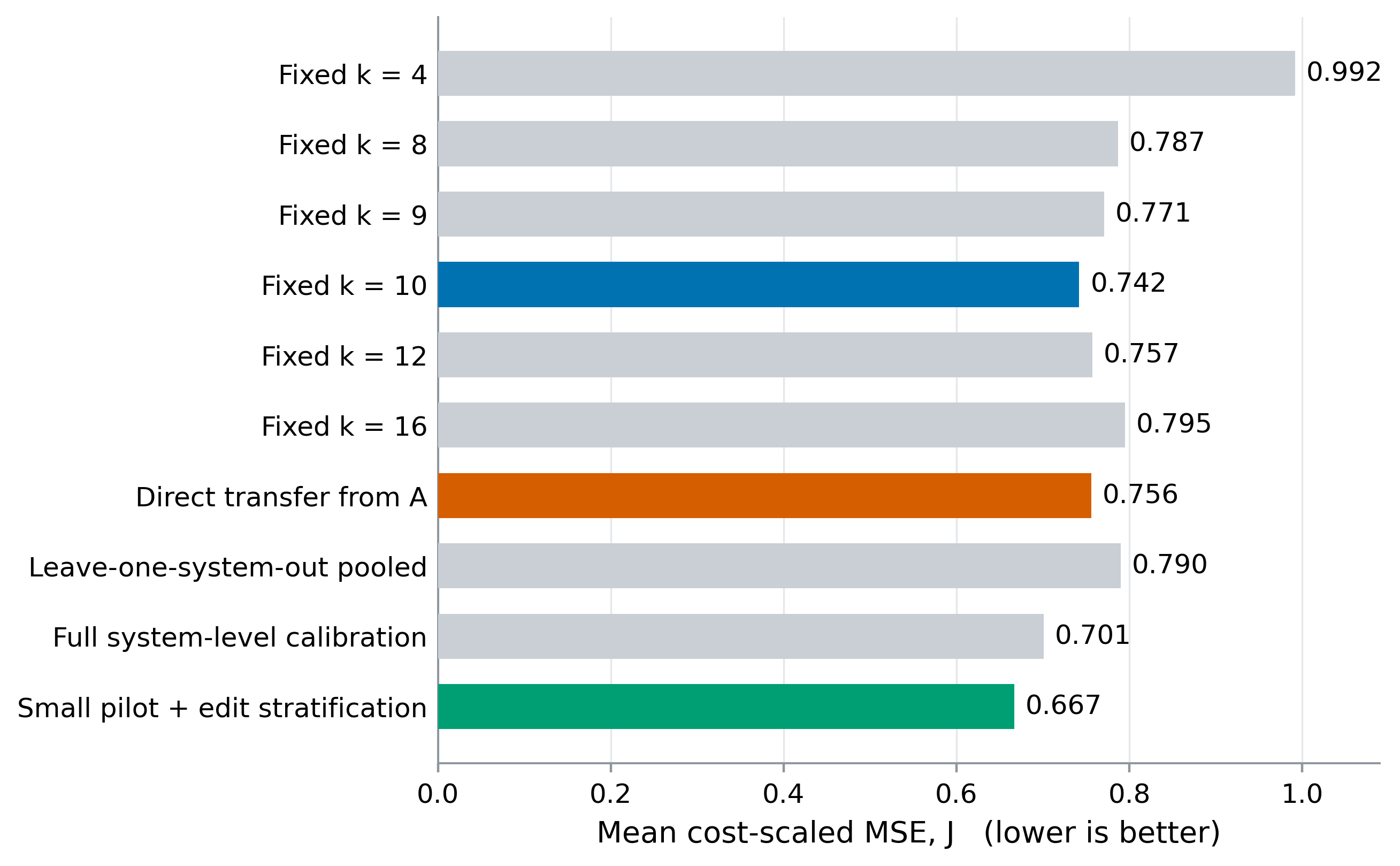}
\caption{Mean cost-scaled error J for each strategy in
cross-task-family environment B; lower is better. Blue highlights the
best reported fixed depth, k=10; orange denotes direct transfer from A
with g=1, k=12; green denotes small calibration with edit
stratification; other strategies are gray. The formal budget is \(C_0\)=1024,
with additional calibration cost reported separately. Exact values are
reported in Table~\ref{tab:F8}.}
\label{fig:3}
\end{figure}

The second CAD environment also exhibits different configuration
preferences. Among fixed-depth baselines, \(k = 10\) has the lowest mean
\(J\) at 0.742. The pooled configuration transferred directly from the
primary environment keeps \(g = 1,k = 12\), with mean \(J = 0.756\).
Leave-one-system-out pooling has mean \(J\) of 0.790, while small
calibration with four-way stratification achieves the lowest mean in the
table, \(J = 0.667\). Configurations are compared in Figure~\ref{fig:3}.

Cross-task-family transfer shows that the primary-environment configuration can directly transfer its frozen \(g,k\), but optimality is not guaranteed on a new task set. Changes in task composition alter both variance structure and execution cost, so transfer still requires remeasuring costs and comparing the transferred configuration against fixed-depth and calibrated strategies in the target environment.

\subsection{Judge Validity and Model Decisions}\label{judge-validity-and-model-decisions}

Dual-expert validation shows that, under inverse-sampling-probability
weighting of the expert sample, the automated judge and expert consensus
preserve the same risk ranking across all five systems. The automated
judge underestimates risk for S4 and S5 by about 1.97 and 3.65
percentage points, respectively, while S2\textquotesingle s automated
labels exactly match expert consensus. Complete weighted risks and
classification metrics are reported in Table~\ref{tab:E1}, with the
validation protocol in Appendix~\ref{appendix-e.-expert-validation}.

Programs that fail the initial build contribute 16 failed states to the
primary risk. We also report conditional edit risk restricted to
initially valid programs. This dual reporting preserves end-to-end
failure probability while enabling analysis of parameter editability
after a program builds successfully.

In model-pair experiments, paired allocation provides only a small
improvement over a strong fixed-depth baseline and the benefit largely
disappears at high budgets. Its value is concentrated in comparisons
with small risk gaps, low paired variance, and reusable calibration
cost. Complete correct-decision rates at three budgets are reported in
Table~\ref{tab:F5}.

\section{Discussion}\label{discussion}
Reliable evaluation requires matching independent evidence to the dominant uncertainty source. Template heterogeneity favors additional tasks, generation variability favors independent programs, and substantial edit-state variation with expensive generation favors deeper auditing. Joint allocation explains system-specific deviations and audit-benefit directions, but does not consistently outperform strong fixed-depth baselines. Calibration is useful only when its gains and reuse justify its additional cost. Transfer results also caution against extrapolating a configuration beyond its task composition \citep{torralba2011,geirhos2020}.

The estimand covers 16 frozen states, not a continuous parameter space or open-ended interaction. The variance model assumes a prespecified sampling design and within-level exchangeability; correlated costs, nonrandom timeouts, or adaptive stopping require additional treatment. Small calibration sets introduce parameter uncertainty, and the judge has limited semantic coverage despite expert validation. The tested families, systems, and kernels also constrain generalization. Average failure risk does not establish engineering safety or worst-case correctness. Appendix~\ref{extended-discussion} retains the full discussion of practical use, calibration, transfer, pairing, and limitations \citep{mitchell2019}.

\section{Conclusion}\label{conclusion}

Under a fixed budget, more thorough CAD-program auditing can reduce the reliability of model evaluation by sacrificing task coverage and independent generations. Three-level variance and measured execution costs explain this conflict and predict audit-benefit directions across the tested systems and environments. No depth is universally best, and joint allocation does not consistently outperform strong fixed-depth baselines. Its value lies in diagnosing the dominant uncertainty source and guiding budget adjustment, with calibration cost, task composition, and reuse determining practical efficiency. Appendix~\ref{supplementary-explanation} retains the extended conclusion.

\clearpage
\subsection*{AI Use Statement}\label{ai-use-statement}

ChatGPT and Codex were used to assist with literature search, language
polishing, code editing, translation, and related writing tasks. All
AI-assisted content, including factual statements, analyses, and
conclusions, was independently reviewed and verified by the authors. The
authors take full responsibility for the accuracy and integrity of the
manuscript.

\bibliographystyle{iclr2027_conference}
\bibliography{references}
\clearpage
\appendix
\numberwithin{equation}{section}

\section{Statistical Derivations and Cost Allocation}\label{appendix-a.-statistical-derivations-and-cost-allocation}

This appendix corresponds to Sections \ref{three-level-variance-decomposition}--\ref{calibration-cost-and-usage-strategy}. The main text retains
only the equations needed to understand the method; here we provide the
design-based derivation, variance-component estimators, continuous
approximation, and calibration break-even condition.

\subsection{Design-Based Derivation of the Three-Level Variance}\label{a.1-design-based-derivation-of-the-three-level-variance}

Let \(Y_{tie}\in\{0,1\}\) denote the failure label for template t, independent
generation i, and frozen edit state e. Each program has M pre-frozen
states, from which k are sampled without replacement; each template has
g independent generations, and a formal evaluation contains q templates.
Program risk is the mean over the M states, and the model-risk estimator
is the average of the sampled template, generation, and state means.

For a fixed program (t,i), the conditional variance under simple random
sampling without replacement is

\begin{equation}\operatorname{Var}(\bar Y_{ti}(k)\mid t,i)=(1/k-1/M)S^2_{E,ti}.\end{equation}

Averaging g independent generations within the same template reduces the
generation- and edit-level contributions by 1/g; averaging over q
templates then gives the superpopulation design variance:

\begin{equation}\operatorname{Var}(\widehat R)=\sigma_T^2/q+\sigma_I^2/(qg)+\frac{1}{qg}(1/k-1/M)\sigma_E^2.\end{equation}

Here, \(\sigma_T^2\) denotes upper-level heterogeneity among template means,
\(\sigma_I^2\) the generation heterogeneity among full-program risks within a
template, and \(\sigma_E^2\) the mean finite-population variance among frozen
states within a program. The decomposition directly shows that
increasing k affects only the third term; uncertainty at the template
and generation levels can be reduced only by increasing q or g.

When the validation target is a frozen reference pool containing Q
templates and G complete generations per template, templates and
generations are also sampled without replacement, yielding the
finite-population form:

\begin{equation}\operatorname{Var}_F(\widehat R)=(1/q-1/Q)\sigma_T^2+\frac1q(1/g-1/G)\sigma_I^2+\frac1{qg}(1/k-1/M)\sigma_E^2.\end{equation}

Equation~\eqref{eq:5} is used only for method validation on the frozen test pool.
Intervals for the future task distribution still use templates as the
outermost cluster and do not treat the finite test-template set as the
complete task population.

\subsection{Method-of-Moments Estimation of Variance Components}\label{a.2-method-of-moments-estimation-of-variance-components}

Full calibration records contain all M=16 states for every program.
Estimation follows the three-level nested structure and explicitly
subtracts lower-level measurement noise so that edit-sampling variation
is not misattributed to generation heterogeneity, nor generation
variation to template heterogeneity.

\begingroup
\setlength{\tabcolsep}{3pt}
\begin{table}[htbp]
\centering
\caption{Estimation order for the three variance components}\label{tab:A1}
\begin{tabular}{@{}
  >{\raggedright\arraybackslash}p{(\columnwidth - 4\tabcolsep) * \real{0.1600}}
  >{\raggedright\arraybackslash}p{(\columnwidth - 4\tabcolsep) * \real{0.4000}}
  >{\raggedright\arraybackslash}p{(\columnwidth - 4\tabcolsep) * \real{0.4400}}@{}}
\toprule
\textbf{Component} & \textbf{Calibration quantity} & \textbf{Implementation detail} \\
\midrule
\(\sigma_E^2\) & Sample variance over the 16 states within each program, averaged
across programs & Preserves behavioral differences within a program \\
\(\sigma_I^2\) & Variance of complete-program risk within a template & If partial
states are used, first subtract the (1/k\ensuremath{-}{}1/M)\(S_E^2\) measurement term \\
\(\sigma_T^2\) & Variance of complete template means & Subtract \(\sigma_I^2\)/g from
finite generations per template; truncate negative residuals to 0 \\
\bottomrule
\end{tabular}
\end{table}
\endgroup

Negative method-of-moments residuals can arise only from finite-sample
noise. The executable protocol truncates them at zero, ensuring that all
three variance components entering budget optimization are finite and
nonnegative. Finite-population estimates on the frozen reference pool
instead use complete template means, complete program means, and
within-program state variances directly.

\subsection{Cost Constraint, Continuous Approximation, and Integer Search}\label{a.3-cost-constraint-continuous-approximation-and-integer-search}

Let \(c_{T}\), \(c_{I}\), and \(c_{E}\) denote the costs of preparing a new template,
generating/parsing/initially building one program, and executing/judging
one edit, respectively. The formal cost of a balanced design is

\begin{equation}C(q,g,k)=q[c_T+g(c_I+kc_E)].\end{equation}

Integer search is the final allocation rule used in this paper. It
enumerates feasible (g,k), purchases the largest q allowed by the budget
for each pair, and selects the minimum finite-population design
variance. This naturally handles template saturation, finite-population
corrections, and unused integer budget.

\begin{quote}
for g = 1,\ldots,G:

for k = 1,\ldots,M:

d = \(c_{T}\) + g(\(c_{I}\) + k \(c_{E}\))

q = min(Q, floor(\(C_0\) / d))

if q \ensuremath{\geq}{} 1: evaluate \(\operatorname{Var}_F(\widehat R)\)

return feasible (q,g,k) with minimum variance
\end{quote}

The continuous relaxation is used only to explain when increasing audit
depth is worthwhile. Holding g fixed and defining d(k)=\(c_{T}\)+g(\(c_{I}\)+k
\(c_{E}\)), substitution of q\ensuremath{\approx}{}\(C_0\)/d(k) gives

\begin{equation}\begin{aligned}V(k\mid g)&\approx\frac{d(k)}{C_0}\left[H_g+\frac{\sigma_E^2}{gk}\right],\\H_g&=\sigma_T^2+\frac{\sigma_I^2}{g}-\frac{\sigma_E^2}{gM}.\end{aligned}\end{equation}

When \(H_{g}\)\textgreater0, the interior stationary point satisfies

\begin{equation}k^*=\sqrt{\frac{\sigma_E^2(c_T+gc_I)}{g^2c_EH_g}}.\end{equation}

This expression is used only for mechanism interpretation. Larger
within-program variance \(\sigma_E^2\) or larger fixed generation cost favors
deeper auditing, while larger template/generation variance or larger
per-edit cost shifts budget toward new templates or generations. All
reported results use the integer search.

\subsection{Calibration Cost and Break-Even Condition}\label{a.4-calibration-cost-and-break-even-condition}

Let \(J_{\mathrm{global}}\) and \(J_{\mathrm{pilot}}\) denote the cost-scaled MSE of the shared rule
and the system-level calibrated rule under formal budget \(C_0\), and let
\(C_{\mathrm{pilot}}\) be the additional cost of one calibration. If the same
calibration can be reused across multiple formal evaluations, the
continuous approximation to the break-even number of evaluations is

\begin{equation}N_{\mathrm{break}}\approx\frac{C_{\mathrm{pilot}}}{C_0[1-J_{\mathrm{pilot}}/J_{\mathrm{global}}]},\qquad J_{\mathrm{pilot}}<J_{\mathrm{global}}.\end{equation}

If \(J_{\mathrm{pilot}}\)\ensuremath{\geq}{}\(J_{\mathrm{global}}\), no positive break-even count exists. This
approximation is used only to assess whether separate calibration for
one system is worthwhile. We therefore report calibration cost
separately from the formal evaluation budget and treat full calibration
as an analytical reference rather than a default deployment step.

\section{Task Families and Construction of Frozen Edit States}\label{appendix-b.-task-families-and-construction-of-frozen-edit-states}

This appendix corresponds to Sections \ref{data-and-evaluation-environments}--\ref{counterfactual-edit-states-and-automated-judge}. State construction reads
only the task specification; candidate-program implementation, model
identity, and runtime outcomes never enter the state-generation process.

\subsection{Fourteen Task Families and Their Edit Semantics}\label{b.1-fourteen-task-families-and-their-edit-semantics}

\begingroup
\setlength{\tabcolsep}{3pt}
\begin{longtable}[]{@{}
  >{\raggedright\arraybackslash}p{(\columnwidth - 8\tabcolsep) * \real{0.0645}}
  >{\raggedright\arraybackslash}p{(\columnwidth - 8\tabcolsep) * \real{0.1898}}
  >{\raggedright\arraybackslash}p{(\columnwidth - 8\tabcolsep) * \real{0.1687}}
  >{\raggedright\arraybackslash}p{(\columnwidth - 8\tabcolsep) * \real{0.2038}}
  >{\raggedright\arraybackslash}p{(\columnwidth - 8\tabcolsep) * \real{0.3732}}@{}}
\caption{Task families, linked parameters, and high-level semantic contracts}\label{tab:B1}\\
\toprule\noalign{}
\textbf{Env.} & \textbf{Task families} & \textbf{Primary} & \textbf{Linked} & \textbf{Semantic contract} \\
\midrule\noalign{}
\endfirsthead
\toprule
\textbf{Env.} & \textbf{Task families} & \textbf{Primary} & \textbf{Linked} & \textbf{Semantic contract} \\
\midrule\noalign{}
\endhead
\bottomrule\noalign{}
\endlastfoot
A & \path{mounting_bracket} & length & length + width & hole spacing;
symmetric hole array; upright connected to base \\
A & \path{flange_plate} & outer diameter & outer diameter + bolt circle & bolt
circle; equally spaced holes; central bore concentric \\
A & \path{stepped_shaft} & shaft length & shaft + shoulder diameter & shoulder
length; wider shoulder; concentric axial bore \\
A & \path{electronics_enclosure} & length & length + width & wall; interior
clearance follows the outer profile; preserve a single shell \\
A & \path{belt_pulley} & outer diameter & outer + hub diameter & hub length;
concentric hub/rim; through bore \\
A & \path{ribbed_angle} & length & width + height & rib count; uniformly
distributed ribs; base/web connected \\
A & \path{bolt_pattern_plate} & length & spacing x + spacing y & hole
diameter; four-hole biaxial symmetry; holes remain within the plate \\
A & \path{pipe_clamp} & outer diameter & outer + inner diameter & lug length;
concentric pipe bore; mirrored dual lugs \\
B & \path{gear_blank} & outer diameter & outer + root diameter & teeth;
radially uniform teeth; concentric bore \\
B & hinge & leaf length & leaf width + barrel diameter & knuckle count;
cover the hinge axis; continuous pin bore \\
B & \path{drawer_handle} & span & span + mount spacing & grip height;
symmetric mounts; grip spans the two bosses \\
B & \path{bottle_cap} & outer diameter & outer diameter + height & rib count;
cap remains hollow; uniformly spaced exterior ribs \\
B & \path{lattice_panel} & length & length + width & bar count; closed frame
on all four sides; uniformly spaced bars \\
B & \path{bearing_housing} & outer diameter & base length + base width & bore
diameter; concentric bore/seat; connected base; four-hole symmetry \\
\end{longtable}
\endgroup

Environment A contains eight task families with nine variants each;
environment B contains six disjoint task families with eight variants
each. Frozen rules vary multiple dimensional ratios across variants, and
variants are not allowed to degenerate into uniformly scaled copies of
the same geometry.

\subsection{Construction and Acceptance Criteria for the Four State Types}\label{b.2-construction-and-acceptance-criteria-for-the-four-state-types}

\begingroup
\setlength{\tabcolsep}{3pt}
\begin{longtable}[]{@{}
  >{\raggedright\arraybackslash}p{(\columnwidth - 6\tabcolsep) * \real{0.0985}}
  >{\raggedright\arraybackslash}p{(\columnwidth - 6\tabcolsep) * \real{0.2214}}
  >{\raggedright\arraybackslash}p{(\columnwidth - 6\tabcolsep) * \real{0.3414}}
  >{\raggedright\arraybackslash}p{(\columnwidth - 6\tabcolsep) * \real{0.3387}}@{}}
\caption{Construction contract for frozen edit states}\label{tab:B2}\\
\toprule\noalign{}
\textbf{Type} & \textbf{Construction target} & \textbf{Acceptance criteria} & \textbf{Primary stress} \\
\midrule\noalign{}
\endfirsthead
\toprule
\textbf{Type} & \textbf{Construction target} & \textbf{Acceptance criteria} & \textbf{Primary stress} \\
\midrule\noalign{}
\endhead
\bottomrule\noalign{}
\endlastfoot
local & Routine single-parameter edit & At least one parameter changes;
full vector is valid; normalized distance \ensuremath{\geq}{}0.10 & Parameter binding and
local geometric response \\
boundary & Near a true feasible boundary & Activates an engineering
constraint; normalized constraint margin \ensuremath{\leq}{}0.05 & Thin walls, spacing,
containment, and degenerate geometry \\
linked & Predefined related-parameter group & At least two linked
parameters change simultaneously; complete dependency closure &
Dimensional relations, symmetry, and multi-parameter consistency \\
semantic & High-level semantic parameters and contract & dependency
closure \ensuremath{\geq}{}2; semantic contract exists; geometric signature must respond &
Design semantics such as array count, shell structure, assembly, and
concentricity \\
\end{longtable}
\endgroup

Each template contains exactly four states of each of the four types,
for 16 complete parameter vectors. Construction rejects no-ops,
out-of-range states, and duplicate vectors. Each state records units,
valid ranges, changed parameters, expected relations, and tolerances.
Formal evaluation samples only from these 16 frozen states without
replacement and never adapts the state set after seeing the candidate
program.

\subsection{State-Quality and Split-Consistency Checks}\label{b.3-state-quality-and-split-consistency-checks}

\begingroup
\setlength{\tabcolsep}{3pt}
\begin{longtable}[]{@{}
  >{\raggedright\arraybackslash}p{(\columnwidth - 2\tabcolsep) * \real{0.2078}}
  >{\raggedright\arraybackslash}p{(\columnwidth - 2\tabcolsep) * \real{0.7922}}@{}}
\caption{Executable challenge-quality checks}\label{tab:B3}\\
\toprule\noalign{}
\textbf{Check} & \textbf{Criterion} \\
\midrule\noalign{}
\endfirsthead
\toprule
\textbf{Check} & \textbf{Criterion} \\
\midrule\noalign{}
\endhead
\bottomrule\noalign{}
\endlastfoot
Uniqueness and validity & Each template has 16 unique parameter vectors;
all parameters lie within frozen valid ranges \\
Edit magnitude & Each state has normalized distance \ensuremath{\geq}{}0.10 from
nominal \\
Boundary stress & Must target a true active constraint, with normalized
engineering margin \ensuremath{\leq}{}0.05 \\
Semantic closure & Each semantic state links at least two parameters and
carries a nonempty semantic contract \\
Variant diversity & Variants within a task family cannot all be
uniform-scale copies \\
Split integrity & Calibration/test are deterministically split within
each family; calibration cannot be a simple variant-number prefix \\
\end{longtable}
\endgroup

These checks cover 120 templates and 1,920 frozen states. They ensure
that deeper auditing adds evidence from the same target population
rather than raising pass rates by lowering task difficulty or rewriting
states for candidate programs.

\section{Isolated Execution and Automated Judge}\label{appendix-c.-isolated-execution-and-automated-judge}

This appendix corresponds to Section~\ref{counterfactual-edit-states-and-automated-judge}. Candidate programs must expose
build(params) and return a measurable CadQuery workplane or shape. The
nominal state and every frozen state are executed in a fresh subprocess.

\subsection{Isolated Worker and Geometry Probes}\label{c.1-isolated-worker-and-geometry-probes}

\begingroup
\setlength{\tabcolsep}{3pt}
\begin{longtable}[]{@{}
  >{\raggedright\arraybackslash}p{(\columnwidth - 4\tabcolsep) * \real{0.1416}}
  >{\raggedright\arraybackslash}p{(\columnwidth - 4\tabcolsep) * \real{0.4183}}
  >{\raggedright\arraybackslash}p{(\columnwidth - 4\tabcolsep) * \real{0.4401}}@{}}
\caption{Geometry probes produced by the isolated worker}\label{tab:C1}\\
\toprule\noalign{}
\textbf{Category} & \textbf{Recorded fields} & \textbf{Use} \\
\midrule\noalign{}
\endfirsthead
\toprule
\textbf{Category} & \textbf{Recorded fields} & \textbf{Use} \\
\midrule\noalign{}
\endhead
\bottomrule\noalign{}
\endlastfoot
process & return code; timeout; wall-clock; stdout/stderr; failure stage
& failure attribution \\
shape & volume; surface area; is\_valid; solid/face/edge/vertex counts &
validity and topology \\
spatial & bounding-box bounds/dimensions; center of mass & dimensions;
containment; symmetry \\
features & face types; circular-edge center/radius/axis & holes; axes;
concentricity \\
signature & hash of sorted face-geometry summaries & edit-response
detection \\
\end{longtable}
\endgroup

Infrastructure-level worker errors are retried at most once. Timeouts,
model-code errors, and geometric build failures do not trigger
model-level retries. If the nominal program is invalid because of model,
code, or geometry failure, all 16 counterfactual states count as
failures in the primary risk. Infrastructure errors retain a separate
failure stage and are never rewritten as model failures.

\subsection{Six-Stage Automated-Judgment Contract}\label{c.2-six-stage-automated-judgment-contract}

\begingroup
\setlength{\tabcolsep}{3pt}
\begin{longtable}[]{@{}
  >{\raggedright\arraybackslash}p{(\columnwidth - 4\tabcolsep) * \real{0.1274}}
  >{\raggedright\arraybackslash}p{(\columnwidth - 4\tabcolsep) * \real{0.3446}}
  >{\raggedright\arraybackslash}p{(\columnwidth - 4\tabcolsep) * \real{0.5280}}@{}}
\caption{Ordered checks performed by the automated judge}\label{tab:C2}\\
\toprule\noalign{}
\textbf{Stage} & \textbf{Inputs} & \textbf{Rule} \\
\midrule\noalign{}
\endfirsthead
\toprule
\textbf{Stage} & \textbf{Inputs} & \textbf{Rule} \\
\midrule\noalign{}
\endhead
\bottomrule\noalign{}
\endlastfoot
1 Build & process status; timeout & build finishes within limit;
measurable shape returned \\
2 Entity & is\_valid; solid/component counts & valid entities; required
component count \\
3 Topology & face/edge/circle probes & feature counts, radii and
locations match task contract \\
4 Response & nominal/edit probes; geometry signature & edited parameters
cause the expected detectable response \\
5 Geometry & bbox; centers; spacing; margins & dimensions, symmetry,
containment and clearance pass \\
6 Semantic & expected relations; dependency closure & high-level
semantic contract remains satisfied \\
\end{longtable}
\endgroup

Length tolerance is max(0.05 mm, 0.001\ensuremath{\cdot}{}\(L_{\mathrm{ref}}\)), angular tolerance is
0.1\ensuremath{^\circ}{}, and relative volume tolerance is 0.5\%; discrete topology counts
use exact rules. A 0.90T--1.10T gray zone is frozen around numerical
thresholds. Gray-zone records are routed to manual review and excluded
from analyses requiring a binary outcome until a review label is
available.

\subsection{Audit-Execution Pseudocode}\label{c.3-audit-execution-pseudocode}

\begin{quote}
run nominal build(params\_nominal) in a fresh subprocess

if infrastructure failure: emit infrastructure stage; do not relabel as
model failure

if nominal model/code/geometry failure: assign failure to all 16 frozen
states

else:

cache nominal geometry probes

for each selected frozen state:

execute build(params\_state) in a fresh subprocess

extract probes; apply ordered judge stages 1\ldots6

emit state\_id, edit\_type, label, failure\_stage, timing and
diagnostics
\end{quote}

\section{System Mapping, Runtime Environment, and Frozen Protocol}\label{appendix-d.-system-mapping-runtime-environment-and-frozen-protocol}

This appendix corresponds to Sections \ref{data-and-evaluation-environments} and \ref{data-splits-and-calibration-setup} and summarizes only
the fixed configurations required to reproduce the experiments. External
repository identifiers unrelated to the method or results are omitted
from the anonymous manuscript.

\subsection{Generation Systems}\label{d.1-generation-systems}

\begingroup
\setlength{\tabcolsep}{3pt}
\begin{longtable}[]{@{}
  >{\raggedright\arraybackslash}p{(\columnwidth - 6\tabcolsep) * \real{0.0808}}
  >{\raggedright\arraybackslash}p{(\columnwidth - 6\tabcolsep) * \real{0.3160}}
  >{\raggedright\arraybackslash}p{(\columnwidth - 6\tabcolsep) * \real{0.3019}}
  >{\raggedright\arraybackslash}p{(\columnwidth - 6\tabcolsep) * \real{0.3013}}@{}}
\caption{Five generation systems}\label{tab:D1}\\
\toprule\noalign{}
\textbf{System} & \textbf{Underlying model and fixed version} & \textbf{Generation method} & \textbf{Output interface} \\
\midrule\noalign{}
\endfirsthead
\toprule
\textbf{System} & \textbf{Underlying model and fixed version} & \textbf{Generation method} & \textbf{Output interface} \\
\midrule\noalign{}
\endhead
\bottomrule\noalign{}
\endlastfoot
S1 & GPT-4.1 mini & Independent generation under a common prompt &
CadQuery program \\
S2 & GPT-5.5 & Same as above & Same as above \\
S3 & Gemini 2.5 Flash & Same as above & Same as above \\
S4 & Gemini 3.1 Pro & Same as above & Same as above \\
S5 & Claude Sonnet 4.6 & Same as above & Same as above \\
\end{longtable}
\endgroup

\subsection{Environments and Data Matrix}\label{d.2-environments-and-data-matrix}

\begingroup
\setlength{\tabcolsep}{3pt}
\begin{longtable}[]{@{}
  >{\raggedright\arraybackslash}p{(\columnwidth - 4\tabcolsep) * \real{0.2773}}
  >{\raggedright\arraybackslash}p{(\columnwidth - 4\tabcolsep) * \real{0.3614}}
  >{\raggedright\arraybackslash}p{(\columnwidth - 4\tabcolsep) * \real{0.3614}}@{}}
\caption{Complete pools for the two frozen environments}\label{tab:D2}\\
\toprule\noalign{}
\textbf{Item} & \textbf{Environment A} & \textbf{Environment B} \\
\midrule\noalign{}
\endfirsthead
\toprule
\textbf{Item} & \textbf{Environment A} & \textbf{Environment B} \\
\midrule\noalign{}
\endhead
\bottomrule\noalign{}
\endlastfoot
Task families / templates & 8 / 72 & 6 / 48 \\
calibration / test & 24 / 48 & 12 / 36 \\
Complete generations/template & 5 & 4 \\
Frozen states/program & 16 (4\ensuremath{\times}{}4 categories) & 16 (4\ensuremath{\times}{}4 categories) \\
state-level audit records & 28,800 & 15,360 \\
\end{longtable}
\endgroup

\subsection{Execution, Cost, and Randomness Protocol}\label{d.3-execution-cost-and-randomness-protocol}

\begingroup
\setlength{\tabcolsep}{3pt}
\begin{longtable}[]{@{}
  >{\raggedright\arraybackslash}p{(\columnwidth - 2\tabcolsep) * \real{0.2902}}
  >{\raggedright\arraybackslash}p{(\columnwidth - 2\tabcolsep) * \real{0.7098}}@{}}
\caption{Frozen execution protocol}\label{tab:D3}\\
\toprule\noalign{}
\textbf{Protocol item} & \textbf{Fixed value} \\
\midrule\noalign{}
\endfirsthead
\toprule
\textbf{Protocol item} & \textbf{Fixed value} \\
\midrule\noalign{}
\endhead
\bottomrule\noalign{}
\endlastfoot
Master seed & 20270901 \\
Formal budget & \(C_0\)=1024; sensitivity budgets 512 / 1024 / 2048 \\
Template cost / edit cost & 2 / 1 (normalized to one edit execution) \\
A generation cost, S1--S5 & 1 / 3 / 8 / 16 / 32 \\
B generation cost, S1--S5 & 2 / 6 / 16 / 32 / 64 \\
Software environment & Ubuntu 22.04; Python 3.11; CadQuery 2.5.x; OCCT
7.8.x \\
Resources & 16 vCPU; 64 GB RAM; concurrency=1 \\
timeout & generation 180 s; nominal build 60 s; edit 30 s \\
Retries & infrastructure 1; model/code/geometry 0 \\
Fixed audit depths & 1, 4, 8, 9, 10, 12, 16 \\
Preregistered benefit comparisons & 4\ensuremath{\to}{}8 and 8\ensuremath{\to}{}16 \\
\end{longtable}
\endgroup

Per-run wall-clock time is the raw cost measure. The main estimate uses
a 10\% trimmed mean, with the ordinary mean and median used for
sensitivity checks. Calibration/test splitting always treats the
template as the outermost unit; all generations and state records from
the same template remain in the same split.

\section{Expert Validation}\label{appendix-e.-expert-validation}

This appendix corresponds to Sections \ref{counterfactual-edit-states-and-automated-judge} and \ref{judge-validity-and-model-decisions}. The expert sample is
stratified by system \ensuremath{\times}{} edit type \ensuremath{\times}{} automatic label; each nonempty
stratum contributes 20 records or all available records, for 800 total.
Two experts with parametric-CAD experience independently inspect the
task specification, execution log, and edited geometry without knowing
system identity or automated label, and then form consensus through a
predefined adjudication process.

\subsection{Weighted Metrics and Expert Agreement}\label{e.1-weighted-metrics-and-expert-agreement}

If stratum h contains \(N_{h}\) records in the target audit population and
\(n_{h}\) expert-sampled records, each record receives
inverse-sampling-probability weight \(w_{h}\)=\(N_{h}\)/\(n_{h}\). Expert-reference
risk and automated-judgment risk are both estimated on the same expert
sample using these weights; precision, recall, specificity, and accuracy
use the same weighting. The two risks and weighted confusion metrics in
Table~\ref{tab:E1} therefore share the same target population and weighting
scheme. Cohen\textquotesingle s \ensuremath{\kappa}{} between the two experts is computed
before consensus; the 800 doubly labeled records yield \ensuremath{\kappa}{}\ensuremath{\approx}{}0.797.

\begingroup
\setlength{\tabcolsep}{3pt}
\begin{longtable}[]{@{}
  >{\raggedright\arraybackslash}p{(\columnwidth - 12\tabcolsep) * \real{0.0757}}
  >{\raggedright\arraybackslash}p{(\columnwidth - 12\tabcolsep) * \real{0.1885}}
  >{\raggedright\arraybackslash}p{(\columnwidth - 12\tabcolsep) * \real{0.3182}}
  >{\raggedright\arraybackslash}p{(\columnwidth - 12\tabcolsep) * \real{0.1022}}
  >{\raggedright\arraybackslash}p{(\columnwidth - 12\tabcolsep) * \real{0.1084}}
  >{\raggedright\arraybackslash}p{(\columnwidth - 12\tabcolsep) * \real{0.1022}}
  >{\raggedright\arraybackslash}p{(\columnwidth - 12\tabcolsep) * \real{0.1049}}@{}}
\caption{Expert-reference risk and weighted confusion metrics for the automated judge}\label{tab:E1}\\
\toprule\noalign{}
\textbf{Sys.} & \textbf{Expert risk} & \textbf{Weighted auto. risk} & \textbf{Prec.} & \textbf{Recall} & \textbf{Spec.} & \textbf{Acc.} \\
\midrule\noalign{}
\endfirsthead
\toprule
\textbf{Sys.} & \textbf{Expert risk} & \textbf{Weighted auto. risk} & \textbf{Prec.} & \textbf{Recall} & \textbf{Spec.} & \textbf{Acc.} \\
\midrule\noalign{}
\endhead
\bottomrule\noalign{}
\endlastfoot
S1 & 0.0909 & 0.0957 & 0.950 & 1.000 & 0.995 & 0.995 \\
S2 & \textbf{0.2151} & \textbf{0.2151} & 1.000 & 1.000 & 1.000 &
1.000 \\
S3 & 0.3261 & 0.3188 & 1.000 & 0.977 & 1.000 & 0.993 \\
S4 & 0.4060 & 0.3863 & 1.000 & 0.952 & 1.000 & 0.980 \\
S5 & 0.4825 & 0.4460 & 1.000 & 0.924 & 1.000 & 0.963 \\
\end{longtable}
\endgroup

The main bias of the automated judge comes from missed failures rather
than additional false positives: precision remains 1.000 for S3--S5,
while recall decreases as system difficulty increases. This supports
using automated-judgment risk in the main text as a reviewable
model-level estimate while reporting expert-reference risk alongside it.

\section{Supplementary Experiments, Robustness, and Model Comparison}\label{appendix-f.-supplementary-experiments-robustness-and-model-comparison}

This appendix corresponds to Sections \ref{comparison-methods-and-budgets}--\ref{judge-validity-and-model-decisions}. It contains cost,
stability, and model-comparison results omitted from the main text for
space, together with exact code-level definitions of each strategy.
Exact values for the main-text result figures are collected in Appendix
F.6.

\subsection{Calibration Cost}\label{f.1-calibration-cost}

\begingroup
\setlength{\tabcolsep}{3pt}
\begin{longtable}[]{@{}
  >{\raggedright\arraybackslash}p{(\columnwidth - 6\tabcolsep) * \real{0.1231}}
  >{\raggedright\arraybackslash}p{(\columnwidth - 6\tabcolsep) * \real{0.2615}}
  >{\raggedright\arraybackslash}p{(\columnwidth - 6\tabcolsep) * \real{0.2615}}
  >{\raggedright\arraybackslash}p{(\columnwidth - 6\tabcolsep) * \real{0.3538}}@{}}
\caption{Calibration cost (formal budget \(C_0\)=1024)}\label{tab:F1}\\
\toprule\noalign{}
\textbf{System} & \textbf{Full-calibration cost} & \textbf{Small-calibration cost} & \textbf{Small calibration / formal budget} \\
\midrule\noalign{}
\endfirsthead
\toprule
\textbf{System} & \textbf{Full-calibration cost} & \textbf{Small-calibration cost} & \textbf{Small calibration / formal budget} \\
\midrule\noalign{}
\endhead
\bottomrule\noalign{}
\endlastfoot
S1 & 2088 & 288 & 28.1\% \\
S2 & 2328 & 320 & 31.3\% \\
S3 & 2928 & 400 & 39.1\% \\
S4 & 3888 & 528 & 51.6\% \\
S5 & 5808 & 784 & 76.6\% \\
\end{longtable}
\endgroup

Full calibration serves as a mechanism-analysis reference. Small
calibration is used to study whether system-level configurations can
amortize their additional cost when calibration is reused across
multiple leaderboard rounds or model ablations.

\subsection{Split Stability and Task-Family Extrapolation}\label{f.2-split-stability-and-task-family-extrapolation}

\begingroup
\setlength{\tabcolsep}{3pt}
\begin{longtable}[]{@{}
  >{\raggedright\arraybackslash}p{(\columnwidth - 8\tabcolsep) * \real{0.0827}}
  >{\raggedright\arraybackslash}p{(\columnwidth - 8\tabcolsep) * \real{0.2030}}
  >{\raggedright\arraybackslash}p{(\columnwidth - 8\tabcolsep) * \real{0.2556}}
  >{\raggedright\arraybackslash}p{(\columnwidth - 8\tabcolsep) * \real{0.2030}}
  >{\raggedright\arraybackslash}p{(\columnwidth - 8\tabcolsep) * \real{0.2556}}@{}}
\caption{Random template splits and leave-one-task-family-out validation}\label{tab:F2}\\
\toprule\noalign{}
\textbf{System} & \textbf{Random-split 5\% coverage} & \textbf{Random regret median/P90} & \textbf{LOFO 5\% coverage} & \textbf{LOFO regret median/P90} \\
\midrule\noalign{}
\endfirsthead
\toprule
\textbf{System} & \textbf{Random-split 5\% coverage} & \textbf{Random regret median/P90} & \textbf{LOFO 5\% coverage} & \textbf{LOFO regret median/P90} \\
\midrule\noalign{}
\endhead
\bottomrule\noalign{}
\endlastfoot
S1 & 45.0\% & 1.059 / 1.444 & 25.0\% & 1.286 / 1.709 \\
S2 & 65.0\% & 1.028 / 1.374 & 12.5\% & 1.390 / 1.608 \\
S3 & 62.5\% & 1.042 / 1.161 & 75.0\% & 1.000 / 1.215 \\
S4 & 57.5\% & 1.038 / 1.347 & 37.5\% & 1.083 / 1.210 \\
S5 & 72.5\% & 1.033 / 1.229 & 62.5\% & 1.029 / 1.471 \\
\end{longtable}
\endgroup

Random-split stability is evaluated with 40 family-stratified splits in
environment A: three templates from each task family are used for
calibration and the remaining six for held-out evaluation, giving 200
system \ensuremath{\times}{} split outcomes. LOFO calibrates on seven complete task families
and evaluates on the held-out family, so it measures change in task
composition rather than ordinary random-split noise.

\subsection{Hierarchical Interval Experiment}\label{f.3-hierarchical-interval-experiment}

\begingroup
\setlength{\tabcolsep}{3pt}
\begin{longtable}[]{@{}
  >{\raggedright\arraybackslash}p{(\columnwidth - 6\tabcolsep) * \real{0.1775}}
  >{\raggedright\arraybackslash}p{(\columnwidth - 6\tabcolsep) * \real{0.1754}}
  >{\raggedright\arraybackslash}p{(\columnwidth - 6\tabcolsep) * \real{0.3609}}
  >{\raggedright\arraybackslash}p{(\columnwidth - 6\tabcolsep) * \real{0.2862}}@{}}
\caption{The sole difference between the interval-coverage methods}\label{tab:F3}\\
\toprule\noalign{}
\textbf{Method} & \textbf{outer unit} & \textbf{Variance structure} & \textbf{Critical value / target} \\
\midrule\noalign{}
\endfirsthead
\toprule
\textbf{Method} & \textbf{outer unit} & \textbf{Variance structure} & \textbf{Critical value / target} \\
\midrule\noalign{}
\endhead
\bottomrule\noalign{}
\endlastfoot
Three-level design interval & template & template + generation + edit;
finite-population corrections at each level & \(t_{q-1}\); covers mean risk
of the full frozen test pool \\
program-independent Control & generation program & Treats programs as
independent and does not explicitly retain the template level &
program-level df; uses the same point estimates and samples as the
three-level interval \\
\end{longtable}
\endgroup

The interval experiment fixes q=16, g=3, and k=8 and performs 5,000
replays without replacement for each system. Both methods use exactly
the same sampled records and risk point estimates; coverage differs only
in whether template-level correlation is retained. Section \ref{evidence-hierarchy-and-evaluation-reliability} reports
the coverage results, so the same numbers are not repeated here.

\subsection{Formal Selection Contract for Allocation Strategies}\label{f.4-formal-selection-contract-for-allocation-strategies}

\begingroup
\setlength{\tabcolsep}{3pt}
\begin{table}[htbp]
\centering
\caption{Information boundaries of the allocation strategies}\label{tab:F4}
\begin{tabular}{@{}
  >{\raggedright\arraybackslash}p{(\columnwidth - 6\tabcolsep) * \real{0.1351}}
  >{\raggedright\arraybackslash}p{(\columnwidth - 6\tabcolsep) * \real{0.2621}}
  >{\raggedright\arraybackslash}p{(\columnwidth - 6\tabcolsep) * \real{0.2963}}
  >{\raggedright\arraybackslash}p{(\columnwidth - 6\tabcolsep) * \real{0.3064}}@{}}
\toprule
\textbf{Strategy} & \textbf{Calibration information} & \textbf{Frozen/selected quantities} & \textbf{Test stage} \\
\midrule
fixed depth & Target-system calibration & Fix k; choose g on calibration
& Purchase the maximum q under target-system cost \\
pooled & Multiple calibration systems & Jointly select (g,k) & Map \(q_{s}\)
separately from each system\textquotesingle s cost \\
LOSO pooled & Calibration systems excluding target & Jointly select
(g,k) & Simulate a new system; target provides only cost \\
system-level & Full target-system calibration & Select target-system
(g,k) & Analytical reference, not a default deployment \\
edit-stratified & Four state types from target/pooled calibration &
\(k_{h}\)\ensuremath{\geq}{}1 per type & Equal-weight risk over four types; joint integer
search \\
paired & Calibration keys shared by the model pair & Estimate
three-level variance using \(Z=Y_A-Y_B\) & Compare on the same
template/generation/state keys \\
\bottomrule
\end{tabular}
\end{table}
\endgroup

Fixed-depth and system-level joint search share the same variance
estimates, cost model, budget constraint, and minimization objective.
Fixed-depth search constrains \(k\) to a prespecified depth, while
system-level search jointly searches \(g\) and \(k\). Therefore, when
system-level search selects \(k = k_{0}\), its \(\left( q,g \right)\)
must match the optimal feasible configuration under fixed \(k_{0}\).

The edit term under four-way stratification uses the finite stratified
variance:

\begin{equation}\operatorname{Var}_{E,\mathrm{strat}}=\frac1{qg}\sum_hW_h^2\left(\frac1{k_h}-\frac1{M_h}\right)\sigma_{E,h}^2,\quad W_h=\frac14,\;M_h=4.\end{equation}

For the paired strategy, \(Z_{tie}=Y^A_{tie}-Y^B_{tie}\) is defined on
shared keys, and the same three-level variance and cost constraint are
applied to Z. The oracle enumerates feasible configurations only post
hoc on held-out data; relative regret=\(J_{\mathrm{selected}}\)/\(J_{\mathrm{oracle}}\), and
regret\ensuremath{\leq}{}1.05 defines membership in the 5\% robust region. The oracle
never participates in configuration selection. For cross-environment
transfer, (g,k) calibrated in environment A is frozen and only \(q_{s}\) is
remapped using measured costs in environment B.

\subsection{Paired Model Decisions}\label{f.5-paired-model-decisions}

\begingroup
\setlength{\tabcolsep}{3pt}
\begin{longtable}[]{@{}
  >{\raggedright\arraybackslash}p{(\columnwidth - 6\tabcolsep) * \real{0.4000}}
  >{\raggedright\arraybackslash}p{(\columnwidth - 6\tabcolsep) * \real{0.2000}}
  >{\raggedright\arraybackslash}p{(\columnwidth - 6\tabcolsep) * \real{0.2000}}
  >{\raggedright\arraybackslash}p{(\columnwidth - 6\tabcolsep) * \real{0.2000}}@{}}
\caption{Mean correct-decision rate across ten model pairs}\label{tab:F5}\\
\toprule\noalign{}
\textbf{Method} & \textbf{\(C_0\)=512} & \textbf{\(C_0\)=1024} & \textbf{\(C_0\)=2048} \\
\midrule\noalign{}
\endfirsthead
\toprule
\textbf{Method} & \textbf{\(C_0\)=512} & \textbf{\(C_0\)=1024} & \textbf{\(C_0\)=2048} \\
\midrule\noalign{}
\endhead
\bottomrule\noalign{}
\endlastfoot
Fixed k=8 & 95.8\% & 97.9\% & 99.7\% \\
Fixed k=9 & 95.8\% & 97.7\% & 99.6\% \\
Fixed k=10 & 95.0\% & 98.0\% & 99.7\% \\
Paired system-level allocation & 95.2\% & 98.2\% & 99.6\% \\
Paired system-level + four-way stratification & 95.9\% & 98.4\% &
99.6\% \\
\end{longtable}
\endgroup

The gain from paired allocation is at most sub-percentage-point under
low and medium budgets and largely vanishes at high budget. We therefore
position it in the main text as a focused review tool for closely
matched systems rather than the default configuration for a standard
leaderboard.

\subsection{Exact Data for Main-Text Result Figures}\label{f.6-exact-data-for-main-text-result-figures}

This section reports the complete values, sampling configurations, and
comparison conventions underlying the three result figures in the main
text for ease of lookup and verification.

\begingroup
\setlength{\tabcolsep}{3pt}
\begin{longtable}[]{@{}
  >{\raggedright\arraybackslash}p{(\columnwidth - 8\tabcolsep) * \real{0.0651}}
  >{\raggedright\arraybackslash}p{(\columnwidth - 8\tabcolsep) * \real{0.1692}}
  >{\raggedright\arraybackslash}p{(\columnwidth - 8\tabcolsep) * \real{0.1692}}
  >{\raggedright\arraybackslash}p{(\columnwidth - 8\tabcolsep) * \real{0.1692}}
  >{\raggedright\arraybackslash}p{(\columnwidth - 8\tabcolsep) * \real{0.4274}}@{}}
\caption{Estimation efficiency in the primary environment (corresponding to Figure~\ref{fig:1})}\label{tab:F6}\\
\toprule\noalign{}
\textbf{Sys.} & \textbf{k=4: q/g; J} & \textbf{k=8: q/g; J} & \textbf{k=16: q/g; J} & \textbf{Program-level MSE: k=4\ensuremath{\to}{}8\ensuremath{\to}{}16} \\
\midrule\noalign{}
\endfirsthead
\toprule
\textbf{Sys.} & \textbf{k=4: q/g; J} & \textbf{k=8: q/g; J} & \textbf{k=16: q/g; J} & \textbf{Program-level MSE: k=4\ensuremath{\to}{}8\ensuremath{\to}{}16} \\
\midrule\noalign{}
\endhead
\bottomrule\noalign{}
\endlastfoot
S1 & 46/4; 0.0974 & 48/2; 0.1126 & 48/1; 0.1649 & 0.014284\ensuremath{\to}{}0.004761\ensuremath{\to}{}0 \\
S2 & 44/3; 0.3050 & 42/2; 0.3097 & 48/1; \textbf{0.3014} &
0.026178\ensuremath{\to}{}0.008726\ensuremath{\to}{}0 \\
S3 & 39/2; 0.7178 & 48/1; 0.6533 & 39/1; 0.5080 & 0.038301\ensuremath{\to}{}0.012767\ensuremath{\to}{}0 \\
S4 & 46/1; 1.2881 & 39/1; 0.7728 & 30/1; 0.5378 & 0.043678\ensuremath{\to}{}0.014559\ensuremath{\to}{}0 \\
S5 & 7/4; 2.4026 & 24/1; 1.3358 & 20/1; 0.8429 & 0.045749\ensuremath{\to}{}0.015250\ensuremath{\to}{}0 \\
\end{longtable}
\endgroup

Note: The formal budget is \(C_0\)=1024; J is model-level cost-scaled mean
squared error. See Figure~\ref{fig:1}.

\begingroup
\setlength{\tabcolsep}{3pt}
\begin{longtable}[]{@{}
  >{\raggedright\arraybackslash}p{(\columnwidth - 12\tabcolsep) * \real{0.0672}}
  >{\raggedright\arraybackslash}p{(\columnwidth - 12\tabcolsep) * \real{0.2058}}
  >{\raggedright\arraybackslash}p{(\columnwidth - 12\tabcolsep) * \real{0.1045}}
  >{\raggedright\arraybackslash}p{(\columnwidth - 12\tabcolsep) * \real{0.2058}}
  >{\raggedright\arraybackslash}p{(\columnwidth - 12\tabcolsep) * \real{0.1045}}
  >{\raggedright\arraybackslash}p{(\columnwidth - 12\tabcolsep) * \real{0.2058}}
  >{\raggedright\arraybackslash}p{(\columnwidth - 12\tabcolsep) * \real{0.1062}}@{}}
\caption{Candidate configurations and estimation error under the cost constraint (corresponding to Figure~\ref{fig:2})}\label{tab:F7}\\
\toprule\noalign{}
\textbf{Sys.} & \textbf{Pooled configuration q/g/k} & \textbf{J} & \textbf{System-level candidate q/g/k} & \textbf{J} & \textbf{Stratified candidate q/g/k} & \textbf{J} \\
\midrule\noalign{}
\endfirsthead
\toprule
\textbf{Sys.} & \textbf{Pooled configuration q/g/k} & \textbf{J} & \textbf{System-level candidate q/g/k} & \textbf{J} & \textbf{Stratified candidate q/g/k} & \textbf{J} \\
\midrule\noalign{}
\endhead
\bottomrule\noalign{}
\endlastfoot
S1 & 48/1/12 & 0.1987 & 48/3/5 & 0.1020 & 48/3/5 & 0.0899 \\
S2 & 48/1/12 & 0.3634 & 48/1/16 & 0.3014 & 48/1/16 & 0.3014 \\
S3 & 46/1/12 & 0.4996 & 48/1/11 & 0.5048 & 48/1/11 & 0.4708 \\
S4 & 34/1/12 & 0.6089 & 34/1/12 & 0.6089 & 35/1/11 & 0.5931 \\
S5 & 22/1/12 & 0.9935 & 22/1/12 & 0.9935 & 22/1/12 & 0.9480 \\
\end{longtable}
\endgroup

Note: J=1024\ensuremath{\times}{}MSE; the formal budget is 1024, with calibration cost
reported separately. All configurations use the largest feasible q for
the specified g and k. The stratified configuration samples at least one
state from each category, estimates risk with equal weights across the
four categories, and computes J using the corresponding stratified
variance. Fixed-depth and system-level joint search use the same
variance estimates, cost model, formal budget, and selection objective;
fixed-depth search additionally constrains \(k\) to the specified value.

The table retains absolute J and q/g/k configurations; Figure~\ref{fig:2} plots
each J in this table divided by the Table~\ref{tab:F6} value for the same system
at fixed k=16.

\begingroup
\setlength{\tabcolsep}{3pt}
\begin{longtable}[]{@{}
  >{\raggedright\arraybackslash}p{(\columnwidth - 6\tabcolsep) * \real{0.4968}}
  >{\raggedright\arraybackslash}p{(\columnwidth - 6\tabcolsep) * \real{0.1697}}
  >{\raggedright\arraybackslash}p{(\columnwidth - 6\tabcolsep) * \real{0.1243}}
  >{\raggedright\arraybackslash}p{(\columnwidth - 6\tabcolsep) * \real{0.2091}}@{}}
\caption{Estimation efficiency in the cross-task-family transfer environment (corresponding to Figure~\ref{fig:3})}\label{tab:F8}\\
\toprule\noalign{}
\textbf{Method} & \textbf{Mean \(C_0\)\ensuremath{\times}{}MSE} & \textbf{Mean k} & \textbf{Relative to best in table} \\
\midrule\noalign{}
\endfirsthead
\toprule
\textbf{Method} & \textbf{Mean \(C_0\)\ensuremath{\times}{}MSE} & \textbf{Mean k} & \textbf{Relative to best in table} \\
\midrule\noalign{}
\endhead
\bottomrule\noalign{}
\endlastfoot
Fixed k=4 & 0.992 & 4.0 & +48.8\% \\
Fixed k=8 & 0.787 & 8.0 & +18.0\% \\
Fixed k=9 & 0.771 & 9.0 & +15.6\% \\
Fixed k=10 & 0.742 & 10.0 & +11.4\% \\
Fixed k=12 & 0.757 & 12.0 & +13.5\% \\
Fixed k=16 & 0.795 & 16.0 & +19.3\% \\
\textbf{Direct transfer of pooled configuration from primary
environment} & \textbf{0.756} & \textbf{12.0} & \textbf{+13.4\%} \\
Leave-one-system-out pooled & 0.790 & 11.6 & +18.5\% \\
Full-calibration system-level strategy & 0.701 & 11.4 & +5.1\% \\
Small calibration + four-way stratification & \textbf{0.667} &
\textbf{9.0} & \textbf{0.0\%} \\
\end{longtable}
\endgroup

Note: The formal budget is \(C_0\)=1024, with calibration cost reported
separately. ``Relative to best in table'' uses the lowest mean J in this
table as the reference. Figure~\ref{fig:3} plots the mean J values from this
table.

\section{Anonymous Executable Materials and Consistency Checks}\label{appendix-g.-anonymous-executable-materials-and-consistency-checks}

This appendix lists only the relative paths, analysis entry points, and
validation contracts for the anonymous executable materials accompanying
the paper. It contains no external repository name, username, or URL.

\subsection{Main Artifact Map}\label{g.1-main-artifact-map}

\begingroup
\setlength{\tabcolsep}{3pt}
\begin{longtable}[]{@{}
  >{\raggedright\arraybackslash}p{(\columnwidth - 2\tabcolsep) * \real{0.4203}}
  >{\raggedright\arraybackslash}p{(\columnwidth - 2\tabcolsep) * \real{0.5797}}@{}}
\caption{Executable components in the anonymous supplementary materials}\label{tab:G1}\\
\toprule\noalign{}
\textbf{Path / module} & \textbf{Role} \\
\midrule\noalign{}
\endfirsthead
\toprule
\textbf{Path / module} & \textbf{Role} \\
\midrule\noalign{}
\endhead
\bottomrule\noalign{}
\endlastfoot
\path{data/templates/depthbenchcad_tasks.json} & 120 templates, valid ranges,
split metadata, and 1,920 frozen states \\
\path{data/programs/} & CadQuery reference programs for the templates \\
\path{data/records/*_audits.jsonl} & A/B state-level audit outcomes \\
\path{data/generations/*.jsonl} & generation-level summaries and 16-state
linkage \\
\path{data/records/*expert_annotations.jsonl} & 800 dual-expert validation
records \\
\path{configs/paper_protocol.json} &
seed, budget, cost, timeout, split, depth, replay, expert protocol \\
\path{src/depthbenchcad/audit.py} / executor.py / judge.py & state
construction; isolated execution and probes; six-stage judgment \\
\path{src/depthbenchcad/variance.py} / allocation.py & three-level
variance/FPC; feasible integer allocation \\
\path{src/depthbenchcad/replay.py} / strategies.py & nested
replay; pooled/LOSO/system/stratified/paired \\
\path{src/depthbenchcad/paper_analysis.py} & statistical-analysis layer for
the main text and appendices \\
\path{scripts/reproduce_tables.py} & recompute paper analyses from
record-level inputs \\
\path{scripts/validate_release.py} / validate\_paper\_results.py & release
linkage and checks of paper values/mechanism invariants \\
tests/ & task/state, judge, statistics, strategy, and release tests \\
\end{longtable}
\endgroup

\subsection{Minimal Reproduction Entry Point}\label{g.2-minimal-reproduction-entry-point}

\begin{Verbatim}[fontsize=\small]
python -m pip install -r requirements.txt
python scripts/validate_release.py
pytest -q
python scripts/reproduce_tables.py \
  --records-a data/records/depthbenchcad_A_audits.jsonl \
  --records-b data/records/depthbenchcad_B_audits.jsonl \
  --expert data/records/depthbenchcad_A_expert_annotations.jsonl \
  --out-dir results/reproduced
python scripts/validate_paper_results.py --results results/reproduced
\end{Verbatim}

\subsection{Automated Consistency Checks}\label{g.3-automated-consistency-checks}

\begingroup
\setlength{\tabcolsep}{3pt}
\begin{table}[htbp]
\centering
\caption{Automated checks for the release and statistical analysis}\label{tab:G2}
\begin{tabular}{@{}
  >{\raggedright\arraybackslash}p{(\columnwidth - 2\tabcolsep) * \real{0.1868}}
  >{\raggedright\arraybackslash}p{(\columnwidth - 2\tabcolsep) * \real{0.8132}}@{}}
\toprule
\textbf{Check layer} & \textbf{Frozen invariant} \\
\midrule
corpus & 120 templates; A/B=72/48; 14 families; split=24/48 and 12/36 \\
state population & 16 states/template, 4 per category; unique, valid,
non-no-op; distance\ensuremath{\geq}{}0.10 \\
challenge quality & boundary activates a true constraint; semantic
closure\ensuremath{\geq}{}2; family variants are not uniform-scale copies \\
record linkage & A/B=28,800/15,360 state records; each generation links
to exactly 16 outcomes \\
generation summary & failure\_count, failed\_state\_ids, and edit-type
breakdown match state records exactly \\
statistics & variance components are nonnegative; finite-pool variance
does not exceed the corresponding superpopulation design variance;
integer allocation is cost-feasible \\
expert sample & 800 unique system-template-generation-state keys \\
paper analysis & key table values, benefit directions, regret/coverage,
and frozen mechanism invariants pass an independent validation script \\
\bottomrule
\end{tabular}
\end{table}
\endgroup

These checks do not replace the statistical argument in the paper. Their
purpose is to ensure that task definitions, the state population, record
hierarchy, and analysis inputs contain no silent mismatches, so the
reported results can be recomputed from the frozen record pool under the
same protocol.

\section{Extended Related Work}\label{extended-related-work}

\subsection{Executable CAD Generation and Edit Evaluation}\label{executable-cad-generation-and-edit-evaluation}

CAD generation research has expanded from final geometry to structured
design histories and executable programs. DeepCAD represents CAD
modeling as an operation sequence \citep{wu2021}. Fusion 360 Gallery
provides real design histories and programmatic construction data
\citep{willis2021}. Text2CAD maps natural language to parametric CAD
sequences \citep{khan2024}. CAD-Recode further uses language models
to recover executable CAD code from point clouds \citep{rukhovich2025}. These representations retain parameters, operation order, and
construction logic, enabling generated outputs to remain executable and
editable.

Work toward editable programs spans interactive modeling, program
generation, and reverse engineering. Sketch2CAD converts contextual
sketches into sequential CAD modeling operations \citep{li2020},
while Free2CAD parses freehand drawings into CAD commands \citep{li2022}. ShapeAssembly uses executable programs to describe editable 3D
structures \citep{jones2020}. SECAD-Net recovers sketch-extrude
operations from geometry \citep{li2023}. ComplexGen performs CAD
reconstruction over B-Rep chain complexes \citep{guo2022a}, while
neural halfspace representations learn implicit conversions of manifold
B-Rep solids \citep{guo2022b}. BrepGen uses diffusion models to
generate B-Reps with structured latent geometry \citep{xu2024}.

Related work also obtains compact, interpretable shape representations
through primitives, convex components, or constructive solid geometry.
CSGNet learns to parse constructive solid geometry programs \citep{sharma2018}. BSP-Net generates compact meshes through binary space
partitioning \citep{chen2020}, while CvxNet learns convex
decomposition \citep{deng2020}. CAPRI-Net studies adaptive primitive
assembly for compact CAD shapes \citep{yu2022}. Volumetric primitives
provide compositional shape abstractions \citep{tulsiani2017}, and
superquadric representations extend shape parsing beyond cuboids
\citep{paschalidou2019}. Neural Parts uses invertible neural networks
to learn more expressive 3D part abstractions \citep{paschalidou2021}, while hierarchical cuboid representations support adaptive 3D
shape abstraction \citep{sun2019}.

CAD datasets and representation learning have likewise shifted from
generic 3D geometry toward native engineering structure. ABC provides a
large CAD dataset with explicit parametric surfaces and curves \citep{koch2019}. UV-Net learns directly from B-Rep geometry and topology
\citep{jayaraman2021}, and later work explores self-supervised
representation learning for CAD \citep{jones2023}. AutoMate targets
automatic mating relations in CAD assemblies \citep{jones2021}, while
JoinABLe learns bottom-up assembly of parametric CAD joints \citep{willis2022}. Scan2CAD studies alignment between RGB-D scans and CAD
models \citep{avetisyan2019}, and CAD-Deform adapts CAD models to 3D
scans through deformable fitting \citep{ishimtsev2020}. Together,
these studies show that CAD evaluation must consider geometry, topology,
structural relations, and executability.

In reverse engineering and local geometric understanding, primitive
fitting and structural segmentation provide another geometric foundation
before execution. SPFN performs supervised fitting of parametric
geometric primitives to point clouds \citep{li2019}. ParSeNet fits
parametric surfaces to 3D point clouds \citep{sharma2020}.
PrimitiveNet studies primitive instance segmentation with local
primitive embeddings \citep{huang2021}, while HPNet uses hybrid
representations for primitive segmentation \citep{yan2021}. Point2Cyl
reconstructs editable extrusion cylinders from point clouds \citep{uy2022}. PartNet provides a fine-grained hierarchical benchmark for
part-level 3D understanding \citep{mo2019}. Before learning-based
methods, Efficient RANSAC was widely used for geometric shape detection
in point clouds \citep{schnabel2007}.

Behavioral evaluation also depends on the validity of the automated
judge. For labels that require semantic or expert judgment, automated
metrics alone are insufficient to establish judge reliability.
Cohen\textquotesingle s kappa is a classical measure of agreement for
nominal labels \citep{cohen1960}. A survey in computational linguistics
further emphasizes explicit reporting of annotation protocols,
inter-coder agreement, and the limits of their interpretation \citep{artstein2008}. We therefore combine automated execution logs with
CAD expert review and report precision, recall, expert agreement, and
representative false positives and false negatives by edit type.

\subsection{Budget-Constrained Evaluation and Multistage Sampling}\label{budget-constrained-evaluation-and-multistage-sampling}

Randomness, repeated runs, and statistical power have become central
concerns in reproducible model evaluation. Deep reinforcement learning
experiments show that a single run can obscure substantial stochastic
and implementation variation \citep{henderson2018}. In sequence
labeling, reporting score distributions can alter conclusions about
model differences \citep{reimers2017}. Limited statistical
power also makes both positive and negative conclusions harder to
interpret under finite experimental budgets \citep{card2020}.
Statistical tests for model comparison must match the sampling and
replication design \citep{dietterich1998}. Reusing the same cross-validation
process for model selection and error estimation can introduce
systematic optimism in the final error estimate \citep{varma2006}.

The CAD behavioral evaluation studied here has three levels---task
templates, within-template generations, and finite edit states---and
three cost types: template preparation, program generation, and edit
execution. In stratified sampling, variance and cost at different levels
jointly determine sample allocation \citep{neyman1934}. Under sampling
without replacement, finite populations and inclusion probabilities must
enter the estimator explicitly \citep{horvitz1952}. Repeated
observations within the same template and program also exhibit
within-cluster correlation and cannot be treated as independent \citep{liang1986}. We instantiate these statistical ideas as a
three-level counterfactual audit of executable CAD programs and further
account for execution cost, pilot cost, expert validation, and
task-family transfer.

\section{Extended Discussion and Limitations}\label{extended-discussion}

\subsection{Main Findings and Practical Implications}\label{discussion-findings}

The experiments show that reliable evaluation first requires correctly
identifying independent evidence: ignoring template-level correlation
produces overconfident model conclusions. Deeper fixed auditing is a
strong baseline in the primary environment, while cross-system and
cross-environment differences further show that no single audit depth
dominates universally. The value of audit depth must be interpreted
jointly with the source of variance and execution cost.
Cross-task-family results also caution against extrapolating conclusions
from benchmark-specific distributions; the literature on dataset bias
and shortcut learning shows that an advantage on one benchmark need not
transfer to a new data composition \citep{torralba2011,geirhos2020}.

The main value of the three-level model lies in explaining and
diagnosing evidence allocation and predicting the direction of audit
benefit, without guaranteeing that joint allocation will outperform a
strong fixed-depth baseline in average error. When template variance is
large, more templates should be covered; when within-template generation
variability is large, more independent generations should be sampled;
when edit-state variation is large and program startup cost is high,
deeper within-program auditing is more favorable. The model can
therefore identify when a system or environment departs from a shared
configuration and indicate the appropriate direction of budget
adjustment.

Accounting for calibration cost further narrows the practical scope of
system-level allocation. Full calibration costs more than one formal
evaluation, and the break-even condition for small calibration still
requires further validation. For a new system without target-system
calibration data, the pooled DepthBenchCAD configuration can serve as a
reference point that requires no additional calibration, but the current
results do not support treating it as a universally superior default
relative to fixed depth. System-level calibration is most likely to be
worthwhile when execution costs or variance structure differ
substantially from previously observed systems and enough subsequent
evaluations are expected to amortize the calibration cost.
Model-pair-specific calibration is likewise appropriate for focused
review of closely matched systems and should not be treated as an
implicit cost of a standard leaderboard.

Leave-one-task-family-out and cross-task-family transfer results show
that the preferred depth changes with task composition and its
associated variance and execution costs. The shared configuration
obtained in the primary environment can serve as a reference in similar
settings, but its efficiency should be reported alongside fixed-depth
baselines rather than treated as an optimal constant across systems or
environments. Appendix \ref{appendix-g.-anonymous-executable-materials-and-consistency-checks} lists the reference configurations, analysis
entry points, and consistency checks in the anonymous executable
materials.

The paired experiments support the same conclusion. Relative to a strong
fixed baseline, model-pair-specific allocation yields only limited
average improvement and loses its advantage at high budgets. Pairing
itself can reduce comparison variance; whether additional calibration is
worthwhile depends on model separation, cost shift, and reuse count.

\subsection{Limitations}\label{limitations}

The 16 edit states used here represent only a predefined finite
counterfactual population and cannot cover a continuous parameter space
or open-ended design interaction. The validity of the estimand depends
on the quality of the state generator, valid ranges, and task
constraints. Future work could expand audit coverage through
constraint-driven state generation, importance sampling, and adaptive
stress testing.

The three-level variance formula assumes a prespecified sampling design
and within-level exchangeability. When task difficulty, failure
probability, and execution cost are correlated, or when nonrandom
timeouts and adaptive early stopping occur, design weights or unit-level
cost optimization are needed. Variance components from small calibration
sets may also fluctuate substantially; future work could use REML or
Bayesian hierarchical models to propagate uncertainty in parameter
estimates.

The automated judge reliably detects execution, topology, and explicit
geometric errors, but its coverage of design semantics remains limited.
Expert review can estimate false positives and false negatives but
cannot fully eliminate differences between state definitions and
professional judgment. The current validation also covers a limited
number of task families, candidate systems, and execution environments;
transfer across benchmarks and CAD kernels requires additional real
experiments. Clear reporting of evaluation boundaries, failure modes,
and conditions of use is therefore important for subsequent reuse
\citep{mitchell2019}.

We focus on average failure risk and pairwise risk differences between
models. Engineering safety certification also requires weighting
critical states, worst-case analysis, and continuous-parameter
verification. Unified leaderboards further involve multi-model ranking,
uncertainty propagation, and multiple comparisons. These questions lie
outside the scope of the present estimand.

\section{Supplementary Motivation and Protocol Explanation}\label{supplementary-explanation}

The following paragraphs retain extended motivation, data checks, and interpretation from the full-length manuscript. Core definitions and experimental settings remain in the main text.

Samples in CAD behavioral evaluation are hierarchical: multiple
generations from the same template share task characteristics, and
multiple edits within one program share the same implementation.
Treating these nested observations as independent evidence can
underestimate evaluation uncertainty. Research on stochastic algorithms
has shown that a single run can obscure substantial run-to-run variance
\citep{henderson2018}, while reporting score distributions can reduce
overinterpretation of a single score \citep{reimers2017}. At
the same time, under a fixed budget, adding edit checks to each program
reduces the number of templates or independent generations that can be
covered, creating a tradeoff between thorough program inspection and
accurate model evaluation.

We study an apparently paradoxical question: why can more thorough
auditing of each CAD program make model evaluation less accurate? Under
the same budget, checking every edit state for a small number of
programs may characterize those programs precisely while still failing
to represent the model\textquotesingle s behavior on other tasks and
generations. Allocating part of the budget to new templates or
independent generations can instead yield a more accurate model-level
conclusion. The key is that the three sample types answer different
questions: templates capture task variation, independent generations
capture model stochasticity, and edit states capture behavioral
variation within a program. Statistical tests for model comparison are
highly sensitive to sampling and replication design \citep{dietterich1998}.
Reusing the same data for configuration selection and final evaluation
can also introduce optimistic bias \citep{varma2006}. We therefore
build a three-level evaluation model and ask when the information gained
by increasing one type of evidence is sufficient to offset the loss of
the other two.

We explicitly distinguish task templates (T), model generations
(\(I\mid T\)), and edit states (E), derive a three-level
estimation variance with finite-population corrections, and incorporate
the costs of template preparation, program generation, and edit
execution into a unified budget. The final protocol jointly selects the
number of templates (q), generations per template (g), and audits per
program (k) through integer search.

We further test, on held-out templates and new task families, whether
three-level variance and cost estimates can predict the practical
benefit of deeper auditing. Pilot-study overhead is included to
determine when such allocation is worthwhile in real evaluations.

Across two CAD environments, we compare program-level and model-level
errors at different audit depths and test whether calibration data
predict the direction of the benefit from deeper auditing. Program-level
error can decrease while model-level error increases, and the direction
of the effect across systems is jointly determined by the source of
variance and execution cost. We also test how hierarchical treatment
affects confidence intervals and report calibration overhead separately.

Our analysis builds on classical multistage sampling, with contributions
focused on three-level modeling, finite counterfactual auditing,
end-to-end cost accounting, and systematic empirical validation for
generative CAD. Optimal allocation in stratified sampling traces back to
\citet{neyman1934}. \citet{horvitz1952} formalized estimation without
replacement and inclusion probabilities in finite populations. \citet{liang1986} provided a classical framework for correlated repeated
observations within clusters.

Environment A contains 72 templates, with five generations per template
and 16 edit states per program, yielding 28,800 evaluation records
across five systems. Environment B contains 48 templates, with four
generations per template and 16 states per program, yielding 15,360
records. The calibration/test splits contain 24/48 templates in A and
12/36 templates in B.

The two task sets are checked for near-duplicates before splitting.
Parameter ranges and geometric constraints in environment B are
constructed from its own task specifications, while the judgment
criteria remain consistent with environment A.

Cross-environment validation compares the allocation rule transferred
directly from environment A with a rule recalibrated locally in
environment B; the exact mapping is given in Appendix~\ref{f.4-formal-selection-contract-for-allocation-strategies}.

Equation~\eqref{eq:4} shows that additional checks of the same program can reduce
only edit-state sampling error. Template heterogeneity and generation
stochasticity remain even when all states are inspected. Under a fixed
budget, deeper auditing also reduces the number of templates or
independent generations that can be purchased, so total error may first
decrease and then increase with audit depth. Whether this increase
occurs, and where it begins, depends on the three variance components
and execution costs rather than on the number of edits alone. The
integer search compares these allocations and uses calibration-set
parameters to predict the benefit of deeper auditing on held-out tasks.
The continuous approximation and stationary-point derivation are given
in Appendix~\ref{appendix-a.-statistical-derivations-and-cost-allocation}.

When template or generation differences dominate, deeper auditing
reduces measurement error for an individual program\textquotesingle s
risk but increases total estimation error for model-level risk.
Redirecting the same budget to new templates or independent generations
lowers total error. For systems with greater edit-state variation and
higher generation costs, the relationship reverses and deeper auditing
yields better estimates. The benefit directions predicted from
calibration agree with the held-out observations, showing that
differences in preferred depth across systems can be explained jointly
by evidence source and execution cost.

\subsection{Extended Calibration and Split Protocol}

Variance components and cost parameters are estimated from an
independent calibration set. Generation count and audit depth pooled
across systems serve as a reference configuration that requires no
target-system calibration; the feasible number of templates is then
determined from measured target-system costs. We do not assume that this
configuration outperforms a strong fixed-depth baseline. System-level
calibration is used to analyze system-specific variance and cost
structure and settings in which calibration results can be reused. The
continuous approximation, break-even condition, and paired-calibration
definition appear in Appendices \ref{a.3-cost-constraint-continuous-approximation-and-integer-search} and \ref{f.4-formal-selection-contract-for-allocation-strategies}.

All splits are defined at the template level and stratified within task
family. In the primary environment, each task family contains nine
templates, of which three enter calibration and six enter testing,
yielding 24/48 templates. In the transfer environment, each task family
contains eight templates, with two for calibration and six for testing,
yielding 12/36 templates. All generated programs and edit records from a
template remain in the same split.

The full calibration set executes all 16 states for every program to
estimate three-level variance and execution costs. We also define a
small calibration that samples a few templates per task family to study
shrinkage estimation and reuse scenarios. Sample sizes, costs, and
stability settings are given in Appendices \ref{d.3-execution-cost-and-randomness-protocol} and \ref{f.1-calibration-cost}.

The test set is never used for variance estimation, shrinkage-weight
selection, audit-depth selection, or cost-model fitting. The full test
pool is used only to compute empirical reference risk and to run
resampling replays. This strict separation prevents optimistic error
estimates induced by model or configuration selection \citep{varma2006}.

\subsection{Extended Conclusion}

We study what evidence is sufficient to support reliable conclusions
about generative CAD models under a fixed execution budget. For three
sources of uncertainty---task templates, independent generations from
the same template, and within-program counterfactual edits---we define a
three-level estimator of average failure risk and its finite-population
variance, and place template scheduling, program generation, state
execution, and calibration costs within a unified integer-allocation
framework. The experimental protocol further includes cross-system
pooling, leave-one-system-out calibration, small pilots, edit-type
stratification, and model-pair-specific allocation.

The experiments show that no audit depth is universally best across
systems and environments. Deeper fixed auditing is a strong baseline in
the primary environment, while the optimal fixed depth changes in the
cross-task-family environment. Three-level variance and execution costs
estimated during calibration explain these differences and reliably
predict the direction of the benefit from additional auditing, although
joint allocation strategies do not consistently outperform strong
fixed-depth baselines. The value of system-level and stratified
configurations therefore lies mainly in diagnosing system-specific
deviations and guiding budget adjustment; their net efficiency also
depends on calibration cost, task composition, and reuse count.

Inspecting one program more thoroughly and judging a model more
accurately are objectives at different levels. Under a fixed budget,
they can conflict: additional edit checks reduce within-program
uncertainty while consuming samples that could reveal task heterogeneity
and generation stochasticity. Three-level variance and execution cost
jointly explain this conflict and support prediction of audit benefit.
CAD behavioral evaluation can therefore use the dominant source of
uncertainty to decide when to keep auditing existing programs and when
to spend the next unit of budget on new tasks or generations.

\end{document}